\documentclass[aps,prb,reprint,groupedaddress,showpacs]{revtex4-1}

\usepackage{amsfonts,amssymb,amsmath}

\usepackage{amstext}
\usepackage{graphicx}

\usepackage{paralist}

\usepackage{multirow}
\usepackage{hhline}

\newcommand{\BVMSW}{BVMSW}
\newcommand{\DE}{Damon--Eshbach}

\newcommand{\Or}{\mathcal{O}}

\DeclareMathOperator{\sign}{sign}
\DeclareMathOperator{\Det}{Det}
\DeclareMathOperator{\Imexpl}{Im}

\newcommand{\secref}[1]{Sec.~\ref{#1}}
\newcommand{\Secref}[1]{Section~\ref{#1}}

\newcommand{\equaref}[1]{Eq.~\eqref{#1}}
\newcommand{\Equaref}[1]{Equation~\eqref{#1}}
\newcommand{\equarefs}[1]{Eqs.~\eqref{#1}}
\newcommand{\Equarefs}[1]{Equations~\eqref{#1}}
\newcommand{\explcite}[1]{Ref.~[\onlinecite{#1}]}

\newcommand{\explcites}[1]{Refs.~[\onlinecite{#1}]}
\newcommand{\Explcites}[1]{References~[\onlinecite{#1}]}
\newcommand{\explcitepart}[1]{[\onlinecite{#1}]}

\newcommand{\figref}[1]{Fig.~\ref{#1}}
\newcommand{\Figref}[1]{Figure~\ref{#1}}

\newcommand{\tabref}[1]{Table~\ref{#1}}
\newcommand{\Tabref}[1]{Table~\ref{#1}}
\newcommand{\tabrefs}[1]{Tables~\ref{#1}}

\newcommand{\dd}{\; \mathrm{d}}
\newcommand{\ddnoskip}{\mathrm{d}}

\DeclareMathOperator{\sinc}{sinc}

\begin{document}


\title{Two-dimensional dispersion of magnetostatic volume spin waves}


\author{F. J. Buijnsters}
\email[]{F.Buijnsters@science.ru.nl}
\author{L. J. A. van Tilburg}
\author{A. Fasolino}
\author{M. I. Katsnelson}
\affiliation{Institute for Molecules and Materials, Radboud University, Heyendaalseweg~135, 6525~AJ Nijmegen, Netherlands}


\date{February 3, 2016}

\begin{abstract}
The dipolar (magnetostatic) interaction dominates the behavior of spin waves in magnetic films in the long-wavelength regime.
In an in-plane magnetized film, volume modes exist with a negative group velocity (backward volume magnetostatic spin waves), in addition to the forward surface-localized mode (\DE{}).
Inside the film of finite thickness $L$, the volume modes have a nontrivial spatial dependence, and their two-dimensional dispersion relations $\omega(\mathbf{k})$ can be calculated only numerically.
We present explicit perturbative expressions for the profiles and frequencies of the volume modes, taking into account an in-plane applied field and uniaxial anisotropy, for the regimes $\lVert\mathbf{k}L\rVert \gg 1$ and $\lVert\mathbf{k}L\rVert \ll 1$, which together provide a good indication of the behavior of the modes for arbitrary wavevector $\mathbf{k}$.
Moreover, we derive a very accurate semianalytical expression for the dispersion relation $\omega(\mathbf{k})$ of the lowest-frequency mode that is straightforward to evaluate using standard numerical routines.
Our results are useful to quickly interpret and control the excitation and propagation of spin waves in (opto-)magnetic experiments.
\end{abstract}


\maketitle

\section{Introduction}

The dipolar interaction endows magnetostatic (long-wavelength) spin waves with a very peculiar dynamics.
In an in-plane magnetized ferromagnetic film, their dispersion shows a strong anisotropy originating from the magnetization vector $\mathbf{M}$ \cite{Satoh2012, Hurben1995, Hurben1996}.
Spin waves propagating through the volume of the film appear to move backwards as their group velocity is opposite to their phase velocity (backward volume magnetostatic spin-wave modes, \BVMSW{}) \cite{stancil2009spin}.
Conversely, spin waves with a wavevector $\mathbf{k} \perp \mathbf{M}$ tend to localize near the surface of the film in \DE{} (DE) modes \cite{DamonEshbachPhys}, which are forward modes.
The surface localization of DE modes is exponential, with a decay length inversely proportional to the perpendicular component of $\mathbf{k}$ \cite{stancil2009spin}.

While the backward volume propagation of parallel spin waves $\mathbf{k} \parallel \mathbf{M}$ is well known, in the case of perpendicular propagation $\mathbf{k} \perp \mathbf{M}$ usually only the DE modes are considered \cite{stancil2009spin,Lenk2012,Ciubotaru}.
The DE modes are the most likely to be excited by a microstrip antenna in the \DE{} geometry \cite{DamonEshbachChem} and show unusual features such as nonreciprocal propagation \cite{Jamali2013}.
However, in a film of finite thickness, \BVMSW{}s are not restricted to the case $\mathbf{k} \parallel \mathbf{M}$ and can exist with any (in-plane) wavevector $\mathbf{k}$; in particular, perpendicularly propagating volume modes also exist and have frequencies below the DE branch \cite{Hurben1996,Terui2011}.
The \BVMSW{} modes are, in fact, the dominant modes in optomagnetic \cite{kimel2005ultrafast, kirilyuk2010ultrafast} experiments  as in \explcite{Satoh2012}, where the two-dimensional profile of the initial excitation (almost homogeneous in the film thickness) can be shaped and the subsequent dynamics observed with spatial and temporal resolution.

\begin{figure}
  \includegraphics[scale=1.0]{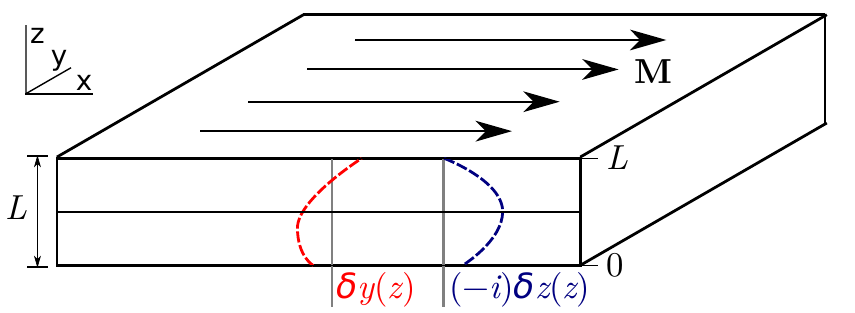}
  \caption{\label{fig:sketch}(color online). Spin-wave modes in ferromagnetic films with a thickness $L$ larger than the exchange length $l$ have a nontrivial perpendicular profile $\delta y(z),\delta z(z)$ inside the film ($0 < z < L$).
   We take $\hat{\mathbf{z}}$ as the film normal and $\hat{\mathbf{x}}$ as the direction of  magnetization $\mathbf{M}$, which is in the plane of the film.
  In the uniform-mode approximation, we assume that $\delta y(z)$ and $\delta z(z)$ are constant functions.
  }
\end{figure}

The propagation of spin waves can only be understood if their dispersion relation is known.
For exchange spin waves (wavelength small compared to the exchange length $l$), as well as for spin waves propagating in an ultrathin film (thin compared to $l$), the dispersion relations are given by fairly simple analytical expressions \cite{Lenk2012}. On micrometer lengthscales, however, exchange interactions are negligible and the film thickness $L$ remains as the only characteristic lengthscale of the system. In this regime, the film can never be considered as effectively two-dimensional, and the perpendicular profile of the volume spin-wave modes, as shown in \figref{fig:sketch}, is essential for an accurate description. 

Because of the nontrivial profile of the mode, the true dispersion relation of the volume modes can in principle be found only numerically \cite{Hurben1996,Satoh2012}.
To our knowledge, the closed-form expressions that have been derived, while useful, rely on either an effectively two-dimensional approach \cite{McMichael2005,Harte1968,Edwards2013,Hurben1998} or on an artificial decoupling of Fourier components \cite{kalinikos1980excitation}.
\Explcites{Hurben1995} and~\explcitepart{DamonEshbachChem}, on the other hand, provide an analytical treatment that is in principle exact but which in practice requires the numerical solution of a set of coupled transcendental equations.

In this article, we study the dispersion and depth profile of \BVMSW{} modes, with a particular focus on the case that $\mathbf{k} \nparallel \mathbf{M}$.
The profiles of such modes show an interesting asymmetry in the perpendicular coordinate~$z$, reminiscent of the asymmetry of the DE modes but without actual surface localization \cite{Hurben1996,kalinikos1980excitation}.
We derive explicit expressions for the mode profiles $\delta y(z),\delta z(z)$, up to first order in $kL$ or $(kL)^{-1}$.
Such expressions allow one, for example, to estimate to what extent the various spin-wave modes couple to an excitation homogeneous in $z$ \cite{Satoh2012}.
We also present, in \tabref{tab:asymp}, simple analytical expressions describing the asymptotic behavior of the dispersion relations of the \BVMSW{} modes.

In addition, we present a practical and very accurate semianalytical approximation to the dispersion relation of the lowest-frequency \BVMSW{} mode, valid on the entire $\mathbf{k}$ plane.
Our expression~\eqref{eq:blockgen} retains the mathematical structure of an eigenvalue equation and is equivalent to the solution of a quartic polynomial equation. It can be evaluated simply and cheaply using standard numerical routines.
We believe that our results are useful for a quick interpretation of experiments and for the development of new applications of directional control of optomagnetic spin-wave excitation \cite{Satoh2012, Lennert2016}.

This article is organized as follows.
In \secref{sec:uniform}, we derive, as a first step, the spin-wave dispersion relation in the uniform-mode (effectively two-dimensional) approximation valid for ultrathin films.
In \secref{sec:genform}, we formulate the normal-mode problem for films of arbitrary thickness. 
In \secref{sec:results}, we describe the typical behavior of the mode profiles and the dispersion relations.
We successfully compare the numerical solutions to our perturbative results, which we present in detail in \secref{sec:limbeh}.
In \secref{sec:semianalytical}, we present our semianalytical expression for the dispersion relation.
\Secref{sec:conclusions} provides a summary of our main conclusions.

\section{\label{sec:uniform}Uniform-mode analysis}

In this section, we review the derivation of the dispersion relation of spin waves in a film in the uniform-mode approximation, where we assume that the precession amplitudes $\delta y(z), \delta z(z)$ of the magnetization inside the film do not depend on the perpendicular coordinate $z$.
Formally, this approximation is valid only in the limit of ultrathin films ($L \ll l$).
While there are some important qualitative differences between the uniform-mode expression and the dispersion relation for large film thickness $L$, it provides a useful first indication of the dispersion behavior of the \BVMSW{} modes.

Specifically, taking $\hat{\mathbf{z}}$ as the film normal, we assume
\begin{equation}\label{eq:unifmode}
\mathbf{M}(t,x,y,z) = M_\text{S}  \mathbf{m}(t,x,y) \Pi^* \Bigl(\frac{z}{L}\Bigr) \text{,}
\end{equation}
where $M_\text{S}$ is saturation magnetization, unit vector $\mathbf{m}(t,x,y)$ is the magnetization direction, and $\Pi^*(z/L)$ is the rectangular function
\begin{equation}
\Pi^*(z/L) =
\begin{cases}
1 &\text{for } 0 < z/L < 1 \\
0 & \text{for } z/L < 0 \text{ or } z/L > 1
\end{cases}
\text{.}
\end{equation}
In view of \secref{sec:regC}, it is convenient to define $\Pi^*(0)=\Pi^*(1)=\tfrac{1}{2}$.

\subsection{Magnetostatic energy: General case}

It is well known that the interaction between two magnetic point dipoles $\mathbf{v}_i,\mathbf{v}_j$ located at $\mathbf{r}_i,\mathbf{r}_j$ is given by
\begin{equation}
E_\text{dip} = - \frac{\mu_0}{4\pi} \frac{3(\mathbf{v}_i\cdot\mathbf{e}_{ij})(\mathbf{v}_j\cdot\mathbf{e}_{ij}) - \mathbf{v}_i\cdot\mathbf{v}_j}{r_{ij}^3}\text{,}
\end{equation}
where $\mathbf{r}_{ij}=\mathbf{r}_{j}-\mathbf{r}_{i}$, $r_{ij}=\lVert\mathbf{r}_{ij}\rVert$, and $\mathbf{e}_{ij}=\mathbf{r}_{ij}/r_{ij}$.
For a continuous magnetization distribution $\mathbf{M}(\mathbf{r}) = M_\text{S} \mathbf{m}(\mathbf{r})$, total energy becomes, in tensor notation,
\begin{align}
\nonumber E_\text{dip}& = \tfrac{1}{2} \mu_0 M_\text{S}^2 \iint m_a(\mathbf{r}') f_{ab}(\mathbf{r}'-\mathbf{r}) m_b(\mathbf{r}) \dd^3 r' \dd^3 r\\
& = \tfrac{1}{2} \mu_0 M_\text{S}^2 \int \tilde{m}_a^*(\mathbf{k}) \tilde{f}_{ab}(\mathbf{k}) \tilde{m}_b(\mathbf{k}) \frac{\ddnoskip^3 k}{(2\pi)^3} \text{,}
\end{align}
where $a,b$ represent the spatial directions $x,y,z$; $\tilde{m}_a(\mathbf{k})$ is the Fourier transform of $m_a(\mathbf{r})$; and where we define
\begin{equation}\label{eq:kernelR}
f_{ab}(\mathbf{r}) = -\frac{A^{(2)}_{ab}(\mathbf{r})}{4\pi r^5} \text{.}
\end{equation}
The factor $1/2$ is a double-counting correction.
The functions $A^{(2)}_{ab}(\mathbf{r})$ are the second-order spherical polynomials
\begin{equation}
A^{(2)}_{ab}(\mathbf{r}) = 3 r_a r_b - \delta_{ab} r_c r_c
\end{equation}
(eg, $A^{(2)}_{xx}(\mathbf{r}) = 3x^2 - r^2$).
The Fourier transform \footnote{
We use the nonunitary definition of the Fourier transform
$\tilde{f}(\mathbf{k}) = \int f(\mathbf{x}) e^{-i\mathbf{k}\cdot\mathbf{x}} \dd^n\mathbf{x}$, where $n$ is the dimension of space.
The inverse transform is given by
$f(\mathbf{x}) = (2\pi)^{-n} \int \tilde{f}(\mathbf{k}) e^{i\mathbf{k}\cdot\mathbf{x}} \dd^n\mathbf{k}$.
We use the result that, for a spherical polynomial $A^{(m)}(\mathbf{r})$ of order $m$, the Fourier transform of a function of the form $f(\mathbf{r}) = f_0(r) A^{(m)}(\mathbf{r})$ is given by
$\tilde{f}(\mathbf{k}) = \bar{f}_0(k) A^{(m)}(\mathbf{k})$,
where
$\bar{f}_0(k) = (2\pi)^{n/2} i^{-m} k^{-(n+2m-2)/2} \int_0^\infty f_0(r) r^{(n+2m)/2} J_{(n+2m-2)/2}(k r)  \dd{}r$ with $J_\alpha(z)$ a Bessel function of the first kind.
} of Eq.~\eqref{eq:kernelR} is given by
\begin{equation}
\tilde{f}_{ab}(\mathbf{k}) = \frac{A^{(2)}_{ab}(\mathbf{k})}{3 k^{2}} \text{.}
\end{equation}

\subsection{Magnetostatic energy: Uniform mode}

For a magnetization profile \eqref{eq:unifmode} that is homogeneous in $z$ inside the film, we have
\begin{align}
\nonumber E_\text{dip}& = \tfrac{1}{2} \mu_0 M_\text{S}^2 \iint m_a(\mathbf{r}') g_{ab}(\mathbf{r}'-\mathbf{r}) m_b(\mathbf{r}) \dd^2 r' \dd^2 r\\
& = \tfrac{1}{2} \mu_0 M_\text{S}^2 L \int \tilde{m}^*_a(\mathbf{k}) \tilde{g}_{ab}(\mathbf{k}) \tilde{m}_b(\mathbf{k}) \frac{\ddnoskip^2k}{(2\pi)^{2}} \text{,}
\end{align}
where
\begin{equation}
g_{ab}(x,y) = \frac{1}{L}  \iint \Pi^*\Bigl(\frac{z'}{L}\Bigr) f_{ab}(x,y,z'-z) \Pi^*\Bigl(\frac{z}{L}\Bigr) \dd{}z'\dd{}z\text{.}
\end{equation}
By the convolution theorem,
\begin{equation}
\tilde{g}_{ab}(k_x,k_y) = L \int_{-\infty}^\infty \tilde{f}_{ab}(k_x,k_y,k_z) \sinc^2 \frac{k_z L}{2} \; \frac{\ddnoskip{}k_z}{2\pi}\text{,}
\end{equation}
where we have used the Fourier transform $\tilde{\Pi}^*(kL) = e^{-ikL/2} \sinc (kL /2)$ with $\sinc\phi = (\sin\phi)/\phi$.
We evaluate
\begin{subequations}\label{eq:greci}
\begin{align}
\tilde{g}_{uv}(k_x,k_y)& = (1-N_k) \frac{k_u k_v}{k^2} - \frac{1}{3}\delta_{uv}\text{,}\\
\tilde{g}_{uz}(k_x,k_y)& = 0\text{,}\\
\tilde{g}_{zz}(k_x,k_y)& = N_k - \frac{1}{3} \text{,}
\end{align}
\end{subequations}
where $u,v$ represent the in-plane coordinates $x,y$.
The demagnetizting factor $N_k$ is given by
\begin{equation}
N_k = \frac{1-e^{-kL}}{kL}\text{.}
\end{equation}
We define $N_{k=0} = 1$ (continuity); notice that $N_{k\rightarrow\infty} = 0$.

Notice that, if we assume that the magnetization of the film is completely homogeneous ($\mathbf{k}=0$), we get $E_\text{dip} = \tfrac{1}{2} \mu_0 M_\text{S}^2 V (-\tfrac{1}{3}m_x^2 - \tfrac{1}{3}m_y^2 + \tfrac{2}{3}m_z^2)$, where $V$ is the total film volume.
Due to the constraint $\lVert\mathbf{m}\rVert=1$, this is effectively a hard-axis anisotropy of strength $\tfrac{1}{2} \mu_0 M_\text{S}^2$, where $\hat{\mathbf{z}}$ is the hard axis. This confirms that the dipolar interaction favors in-plane magnetization, and gives the well-known condition $K>\frac{1}{2}\mu_0 M_\text{S}^2$ for perpendicular (out-of-plane) magnetization due to an intrinsic perpendicular anisotropy $K$ in the absence of an applied field.
As a second limiting case, let us consider a system where $\mathbf{m}(x,y)$ depends only on $x$ ($k_y=0$) and where $L$ is very large (thick film, $L\gg |k_x|^{-1}$). For a fixed $k_x\neq0$, we get, in the limit $L\rightarrow \infty$, an effective local hard-axis anisotropy of strength $\frac{1}{2}\mu_0 M_\text{S}^2$, where the hard axis is $\hat{\mathbf{x}}$.

\subsection{Linearization}

In addition to the dipolar interaction, we take into account the usual micromagnetic energy functionals for exchange $E_\text{ex} = A L \int ( \lVert \partial_x \mathbf{m} \rVert^2 + \lVert \partial_y \mathbf{m} \rVert^2 ) \dd^2r$, intrinsic easy-axis anisotropy $E_\text{ani} = -K L \int m_z^2 \dd^2 r$, and Zeeman energy $E_H = - \mu_0 M_\text{S} H_x L \int m_x \dd^2 r$. The applied field $H_x$ fixes the equilibrium magnetization along $\hat{\mathbf{x}}$.

Linearization of the Landau--Lifshitz equation \cite{Landau1935} without damping 
\begin{equation}
\frac{\partial\mathbf{m}}{\partial t} = \frac{|\gamma|}{M_\text{S} L} \mathbf{m} \times \frac{\delta E}{\delta \mathbf{m}(\mathbf{r})}
\end{equation}
around the equilibrium $\mathbf{m}(\mathbf{r}) = \hat{\mathbf{x}}$
gives, very generally \cite{Buijnsters2014},
\begin{multline}\label{eq:genlin}
\left(
\begin{array}{c c}
 - \frac{\delta E}{\delta x}  &
-\frac{M_\text{S}L}{|\gamma|} \partial_t\\
\frac{M_\text{S}L}{|\gamma|} \partial_t&
 - \frac{\delta E}{\delta x} 
\end{array}
\right)
\left(
\begin{array}{c}
\delta y \\
\delta z
\end{array}
\right)
+ \\
\int
\left(
\begin{array}{c c}
\frac{\delta^2 E}{\delta y\delta y'} &
\frac{\delta^2 E}{\delta y\delta z'}\\
\frac{\delta^2 E}{\delta z\delta y'}&
\frac{\delta^2 E}{\delta z\delta z'}
\end{array}
\right)
\left(
\begin{array}{c}
\delta y' \\
\delta z'
\end{array}
\right)
\dd^2 r'
= 0\text{,}
\end{multline}
where $\gamma$ is the gyromagnetic ratio and
where the functional derivatives of $E$ are to be evaluated for the equilibrium configuration $\mathbf{m}(\mathbf{r}) = \hat{\mathbf{x}}$.
For brevity, we write $\delta y$ for $\delta m_y(t,\mathbf{r})$ and $\delta y'$ for $\delta m_y(t,\mathbf{r'})$.
The functions $\delta y,\delta z$ represent the infinitesimal deviation of magnetization $\mathbf{m}$ from its equilibrium direction.

\begin{figure}
  \includegraphics[scale=1.0]{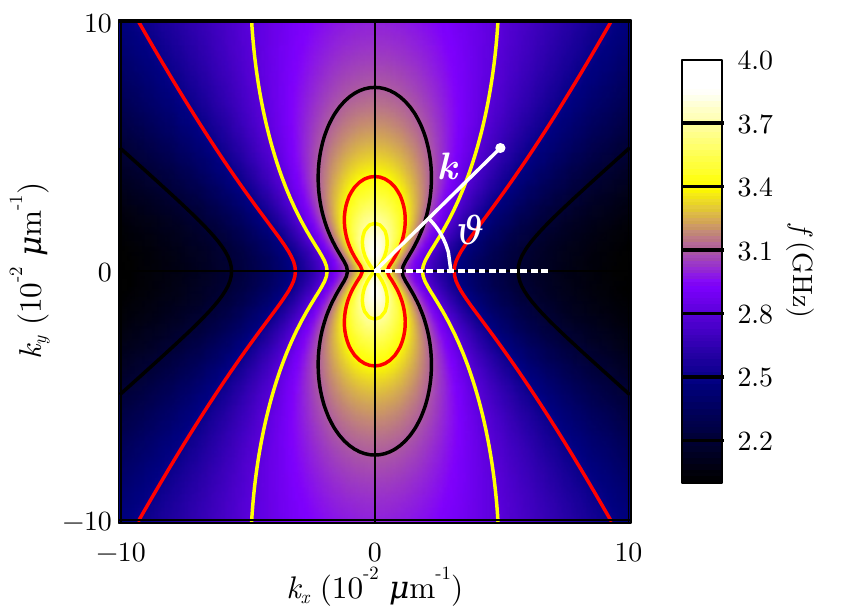}
  \caption{\label{fig:unifmap}(color online). Dispersion relation of BVMSWs in the uniform-mode approximation, for $ H_x = 80\text{ kA/m}$, $K = 3.5\text{ kJ/m}^3$, $L = 100\text{ $\mu$m}$, and $ M_\text{S} = 110\text{ kA/m}$. We neglect exchange $A$, assuming that wavenumber $k$ is much smaller than the inverse exchange length $1/l = \sqrt{\mu_0 M_\text{S}^2/(2A)}$.
  Notice that, in the magnetostatic regime, $\omega$ (mostly) decreases in $k$, giving the spin waves a backward-propagating character.
  We define $\vartheta$ as the polar angle of the wavevector $\mathbf{k} = (k_x,k_y) = (k\cos\vartheta, k\sin\vartheta)$.
  }
\end{figure}

Substituting $E = E_\text{ex} + E_\text{ani} + E_H + E_\text{dip}$ and passing to Fourier space, Eq.~\eqref{eq:genlin} becomes
\begin{multline}\label{eq:unifmodeproblem}
\left(
\begin{array}{c c}
\frac{H_x}{M_\text{S}} + \frac{2 A k^2}{\mu_0 M_\text{S}^2} + (1-N_k) \frac{k_y^2}{k^2} &
-\frac{1}{ \mu_0|\gamma| M_\text{S}} i \omega\\
\frac{1}{\mu_0|\gamma| M_\text{S}} i \omega&
 \frac{H_x}{M_\text{S}} + \frac{2 A k^2 - 2K}{\mu_0 M_\text{S}^2} + N_k
\end{array}
\right) \\
\cdot\left(
\begin{array}{c}
\widetilde{\delta y} \\
\widetilde{\delta z}
\end{array}
\right)
= 0 \text{,}
\end{multline}
The positive solution for $\omega$ in the characteristic equation gives the dispersion relation (\emph{cf.} \explcites{Lenk2012}, \explcitepart{Ciubotaru}, \explcitepart{KalinikosSlavin}, \explcitepart{KittelHerring})
\begin{multline}\label{eq:unifdisp}
\omega =
|\gamma| \mu_0
\sqrt{\Bigl[\frac{2A k^2 - 2K}{\mu_0 M_\text{S}} + H_x + M_\text{S}N_k \Bigr] \times}\\
\overline{\Bigl[\frac{2A k^2}{\mu_0 M_\text{S}} + H_x + M_\text{S} (1-N_k) \sin^2\vartheta\Bigr]}
\text{,}
\end{multline}
where $\vartheta$ is the polar angle of wavevector $\mathbf{k}$, as shown in \figref{fig:unifmap}.

\Figref{fig:unifmap} shows an example of the dispersion relation~\eqref{eq:unifdisp} for typical parameters.
Notice that the dispersion relation has a cusp at the origin $\mathbf{k}=0$.
With the exception of a small area right above and below the point $\mathbf{k}=0$, the frequency decreases with increasing $k$. This implies that the spin waves have a group velocity that is opposite to their wavevector $k$ (backward modes).

\section{\label{sec:genform}General formulation}

We now turn to the general case that film thickness $L$ is not small as compared to exchange length $l$, and the dependence of the modes on the perpendicular coordinate $z$ cannot be neglected. For simplicity, we shall, in fact, assume that both film thickness $L$ and wavelength $2\pi / k$ are much greater than exchange length $l$. This allows us to neglect the exchange energy $E_\text{ex}$.
In the following, whenever we refer to the short-wavelength limit $k\rightarrow\infty$, we mean the regime where the wavelength is much less than film thickness ($kL \gg 1$) but still well above the exchange length $l$ (magnetostatic spin waves, $kl \ll 1$).

Fixing wavenumber $\mathbf{k} = (k_x,k_y)$, we allow the spin-wave mode to have an arbitrary profile $\delta y(z), \delta z(z)$ inside the film.
Analogously to \equaref{eq:unifmodeproblem}, we obtain an eigenvalue equation
\begin{multline}\label{eq:modez}
\left(
\begin{array}{c c}
H_x \hat{S} + M_\text{S} \hat{D}^{yy} & M_\text{S} \hat{D}^{yz} \\
M_\text{S} \hat{D}^{yz} & (H_x - \frac{2K}{\mu_0 M_\text{S}}) \hat{S} + M_\text{S} \hat{D}^{zz} \\
\end{array}
\right)
\left(
\begin{array}{c}
\delta y \\
\delta z \\
\end{array}
\right)
\\
= 
\frac{\omega}{\mu_0 |\gamma|}
\left(
\begin{array}{c c}
0 &  i  \hat{S} \\
- i \hat{S} & 0 \\
\end{array}
\right)
\left(
\begin{array}{c}
\delta y \\
\delta z \\
\end{array}
\right)
\text{,}
\end{multline}
where $\delta y(z),\delta z(z)$ are now functions of $z$, supported on the interval $0\le z \le L$.
Here $\hat{S}$ represents the identity operator. The operators $\hat{D}^{ab}$ may be represented in Fourier space as
\begin{equation}\label{eq:defD}
\hat{D}^{ab}(k_x,k_y) = \frac{k_a k_b}{k_x^2 + k_y^2 + k_z^2}\text{,}
\end{equation}
where $k_x,k_y$ should be treated as numerical constants (parameters of $\hat{D}^{ab}$) but $k_z$ as an operator $\hat{k}_z = -i\partial_z$ acting on the functions $\delta y(z), \delta z(z)$.

The functions $\delta y(z), \delta z(z)$ vanish outside the interval $0 < z < L$. 
The finite film thickness $L$ quantizes the modes that can be excited for any given $k_y,k_z$. We label the modes as $n=1,2,\ldots$ in order of increasing $\omega > 0$.

It is convenient to normalize the solutions $\Psi_+$ to satisfy
\begin{multline}\label{eq:normconstraint}
\Psi_+^\dagger \hat{Q} \Psi_+ =
\frac{1}{ \mu_0 |\gamma|}
\left(
\begin{array}{c}
\delta y \\
\delta z \\
\end{array}
\right)^\dagger 
\left(
\begin{array}{c c}
0 &  i  \hat{S} \\
- i \hat{S} & 0 \\
\end{array}
\right)
\left(
\begin{array}{c}
\delta y \\
\delta z \\
\end{array}
\right)
\\=
\frac{2}{\mu_0 |\gamma|} \Imexpl \int_{0}^{L} \delta y(z) \delta z^*(z)\dd{}z
=
1
\text{,}
\end{multline}
where the asterisk denotes complex conjugation.
Because the cross elements $M_\text{S} \hat{D}^{yz}$ in \equaref{eq:modez} are Hermitian, we may assume without loss of generality that $\delta y(z)$ is purely real and $\delta z(z)$ is purely imaginary.
Notice that if $\Psi_+$ is a solution of \equaref{eq:modez} with eigenvalue~$\omega$, its complex conjugate $\Psi_-$ is a solution with eigenvalue $-\omega$ (and norm $\Psi_-^\dagger \hat{Q} \Psi_- = -1$). The fact that solutions occur in conjugate pairs is a result of the Hamiltonianness of the normal-mode problem \cite{Buijnsters2014}. The negative-$\omega$ solution $\Psi_-$ is redundant.

\subsection{Asymptotic frequencies}

Fixing the polar angle $\vartheta$, we now turn to the behavior of \equaref{eq:modez} in the limits $k\rightarrow0$ and $k\rightarrow\infty$ along a radial half-line $(k_x,k_y) = (k\cos\vartheta, k\sin\vartheta)$.

For $k\rightarrow\infty$, the operator $\hat{D}^{yy}$ reduces to
\begin{equation}
\hat{D}^{yy} = \frac{k_y^2}{k_x^2 + k_y^2 + \hat{k}_z^2} = \frac{k^2\sin^2\vartheta}{k^2 + \hat{k}_z^2 } \rightarrow (\sin^2\vartheta) \hat{S}\text{,}
\end{equation}
which is a simple scalar operator;
analogously, we find
\begin{equation}
\left(
\begin{array}{c c}
\hat{D}^{yy} & \hat{D}^{yz} \\
\hat{D}^{yz} & \hat{D}^{zz} \\
\end{array}
\right)
\rightarrow
\left(
\begin{array}{c c}
(\sin^2\vartheta) \hat{S} & 0 \\
0 & 0 \\
\end{array}
\right)
\text{.}
\end{equation}
We conclude that all modes $n$ are degenerate in the limit $k\rightarrow\infty$, since now only the identity operator $\hat{S}$ acts on $\delta y(z),\delta z(z)$ in \equaref{eq:modez}.
In particular, it follows that the uniform-mode expression~\eqref{eq:unifdisp} for $\omega$ is exact in this limit, and we have \cite{Terui2011}
\begin{equation}\label{eq:freqinf}
\omega_{k\rightarrow\infty} =
|\gamma| \mu_0
\sqrt{
\Bigl(H_x - \frac{2K}{\mu_0 M_\text{S}} \Bigr)
\Bigl(H_x + M_\text{S} \sin^2\vartheta\Bigr)
}
\text{.}
\end{equation}
In the opposite limit $k\rightarrow0$ (uniform precession), we find
\begin{equation}
\left(
\begin{array}{c c}
\hat{D}^{yy} & \hat{D}^{yz} \\
\hat{D}^{yz} & \hat{D}^{zz} \\
\end{array}
\right)
\rightarrow
\left(
\begin{array}{c c}
0 & 0 \\
0 & \hat{S} \\
\end{array}
\right)
\text{,}
\end{equation}
and again all modes $n$ are degenerate; the precession frequency is given by \cite{paper59Kittel,Terui2011}
\begin{equation}\label{eq:freqzero}
\omega_{k=0} =
|\gamma| \mu_0
\sqrt{
\Bigl(H_x + M_\text{S} - \frac{2K}{\mu_0 M_\text{S}} \Bigr) 
H_x
}
\text{,}
\end{equation}
in agreement with the uniform-mode expression~\eqref{eq:unifdisp}.

\subsection{Asymptotic profiles}

In the limits $k\rightarrow0$ and $k\rightarrow\infty$, the only operator acting on the profiles $\delta y(z), \delta z(z)$ is the identity operator $\hat{S}$. The matrices of operators in \equaref{eq:modez} reduce to simple $2\times2$ scalar matrices.
As a result, we can solve \equaref{eq:modez} analytically and multiply the solution vector by an arbitrary function.
We obtain solutions $\Psi_+$ of the form
\begin{equation}\label{eq:asympprof0}
\Psi_0
=
\left(
\begin{array}{c}
\delta y(z) \\
\delta z(z) \\
\end{array}
\right)
=
\sqrt{\frac{\mu_0 |\gamma|}{2 ab}}
\left(
\begin{array}{c}
\phantom{-i}a \, \psi_0(z) \\
-ib \, \psi_0(z)
\end{array}
\right)
\text{,}
\end{equation}
where 
$\psi_0(z)$ is a real-valued function supported on the interval $0 \le z \le L$.
Notice that $\delta y(z), \delta z(z)$ differ only by a scalar factor.
For the regime $k \rightarrow 0$, the values of $a,b$ are given by
\begin{subequations}\label{eq:absmallk}
\begin{align}
a &= \sqrt{H_x + M_\text{S} -\tfrac{2K}{\mu_0 M_\text{S}}} \text{,}\\
b &= \sqrt{H_x}\text{;}
\end{align}
\end{subequations}
for the regime $k \rightarrow \infty$, we have
\begin{subequations}\label{eq:abbigk}
\begin{align}
a &= \sqrt{H_x - \tfrac{2K}{\mu_0 M_\text{S}}} \text{,}\\
b &= \sqrt{H_x + M_\text{S}\sin^2 \vartheta} \text{.}
\end{align}
\end{subequations}
The profile $\psi_0(z)$ must satisfy the normalization condition $\int_{0}^{L}\psi_0(z)^2\dd{}z = 1$, but is otherwise arbitrary.

In \secref{sec:limbeh}, we find that the degeneracy of the modes $n$ is lifted and $\psi_0(z)$ fixed by the higher-order terms in the expansion of \equaref{eq:modez} in $k$ or $k^{-1}$.
For $k\rightarrow0$, the profile $\psi_0(z)$ depends on the angle of approach $\vartheta$ \cite{Hurben1995}.

\section{\label{sec:results}Mode profiles and dispersion}

\begin{figure}
  \includegraphics[scale=1.0]{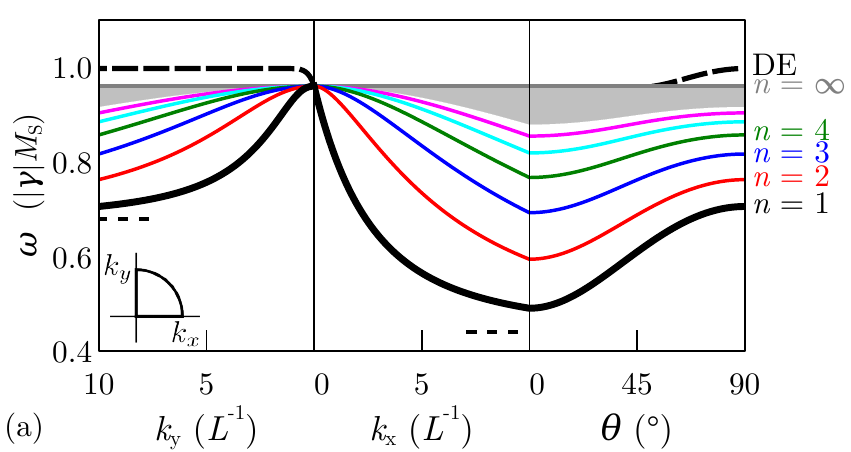}
  \includegraphics[scale=1.0]{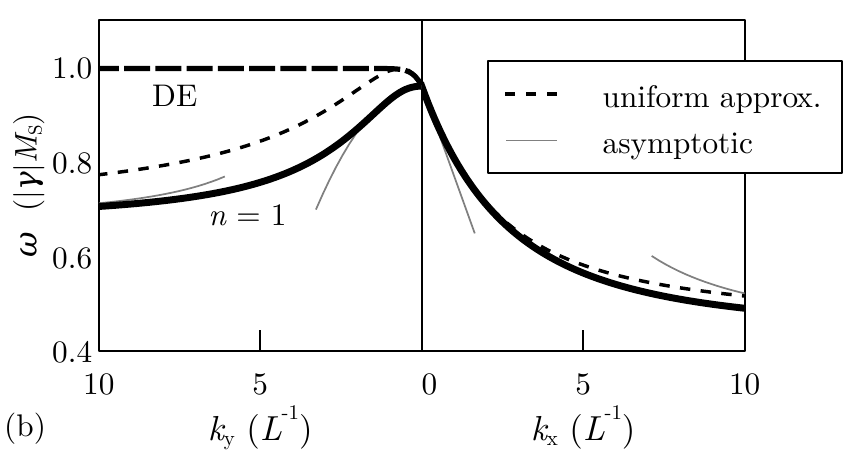}
  \includegraphics[scale=1.0]{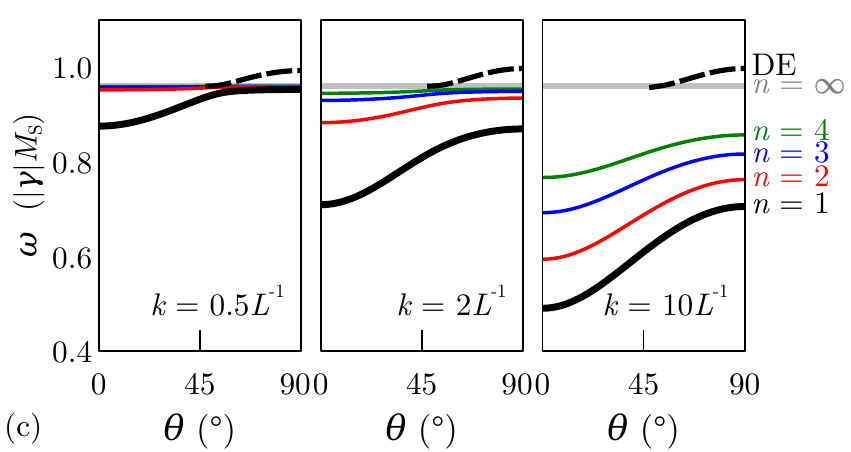}
  \caption{\label{fig:dispavoid}(color online).
  (a)~Numerical dispersion relations of the volume modes $n=1,2,\ldots$ and the DE surface mode, along the $k_y$ axis ($\vartheta=90^\circ$), the $k_x$ axis ($\vartheta=0^\circ$), and a circular arc ($k=10\,L^{-1}$), taking $H_x = 0.73\,M_\text{S}$ and $2K = 0.46\,\mu_0 M_\text{S}^2$.
  For $k=0$, all modes are degenerate, with $\omega = \omega_{k=0}$ given by \equaref{eq:freqzero}.
  Taking the $k\rightarrow\infty$ limit in a fixed direction~$\vartheta$, each volume mode $n$ eventually approaches the frequency $\omega_{k\rightarrow\infty}$ (short dotted lines), given by \equaref{eq:freqinf}.
  (b)~Numerical dispersion relations of the $n=1$ and DE modes, compared to the uniform-mode expression~\eqref{eq:unifdisp}.
  Along the $k_x$ axis, \equaref{eq:unifdisp} is fairly accurate, predicting the correct group velocity $\ddnoskip{}\omega / \ddnoskip k$ of the $n=1$ mode for $k\rightarrow0$. Along the $k_y$ axis, there is a significant deviation.
  The correct asymptotic behavior of the $n=1$ mode (thin solid lines) is given in \tabref{tab:asymp}.
  (c)~For small $k$, the crossover between the $n=1$ and DE modes might be seen as an avoided band crossing.
  The behavior of the $n=1$ mode below $\vartheta_\text{cr}=49^\circ$ is similar to that of the DE mode above $\vartheta_\text{cr}$, but the two modes are not continuously connected; the DE mode instead emerges from the quasicontinuum of high-$n$ volume modes.
  }
\end{figure}

\Figref{fig:dispavoid}(a) shows the dispersion relations obtained from a numerical solution of the eigenvalue equation~\eqref{eq:modez}.
We find a sequence of modes $n=1,2,\ldots$ that can be identified as the \BVMSW{} modes \cite{Hurben1995,Terui2011}. Their frequencies monotonically decrease in any direction $\vartheta$ as we move away from the origin $k=0$.
In addition, we find, for wavevectors $\mathbf{k}$ pointing predominantly along the $k_y$ axis (perpendicular to magnetization), a single special branch, which we identify as the DE surface mode \cite{DamonEshbachPhys}. Its frequency increases in $k$ before leveling off to a constant value.

At $k=0$, all modes have the same frequency $\omega_{k=0}$, given by~\equaref{eq:freqzero}.
The behavior of $\omega(\mathbf{k})$ in the opposite limit $k\rightarrow\infty$ is somewhat more involved.
Any given volume mode $n$ eventually converges to the same frequency $\omega_{k\rightarrow\infty}$, given by \equaref{eq:freqinf}, as we take $k\rightarrow\infty$. However, for any fixed wavevector $\mathbf{k}$, the frequency in the limit $n\rightarrow\infty$ converges to $\omega_{k=0}$. As a consequence, there is a quasicontinuum of high-$n$ volume modes just below the line $\omega = \omega_{k=0}$.

\begin{figure*}
  \includegraphics[scale=1.0]{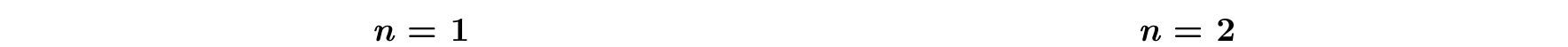}
  \includegraphics[scale=1.0]{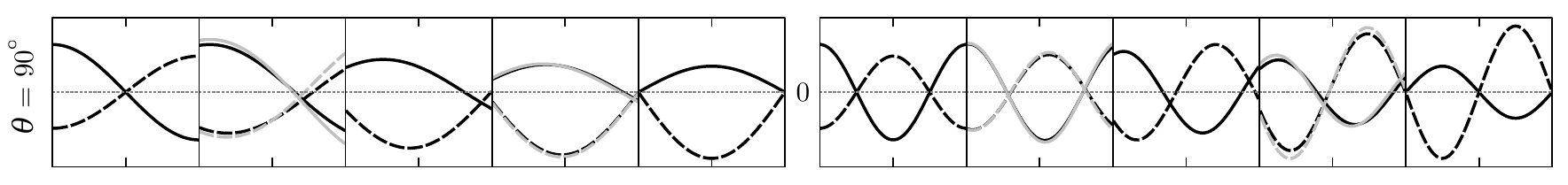}
  \includegraphics[scale=1.0]{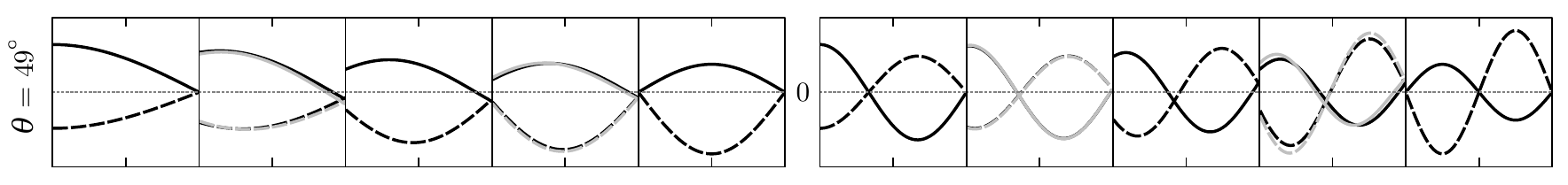}
  \includegraphics[scale=1.0]{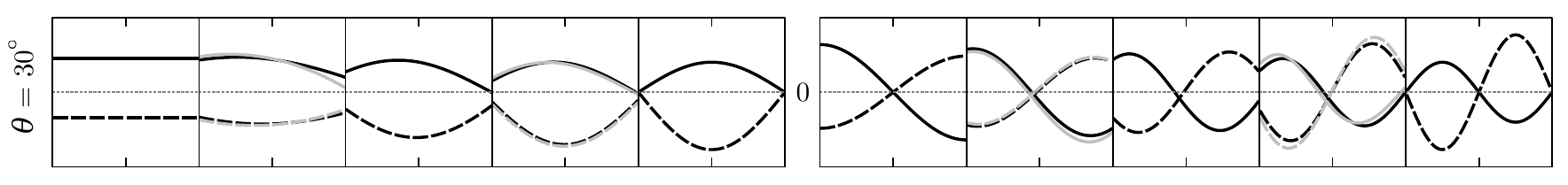}
  \includegraphics[scale=1.0]{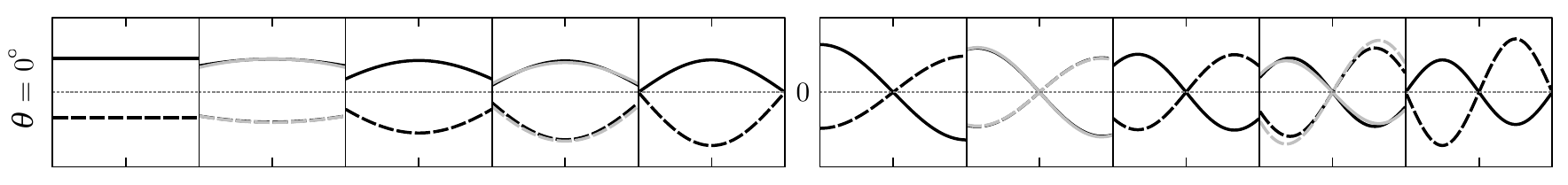}
  \includegraphics[scale=1.0]{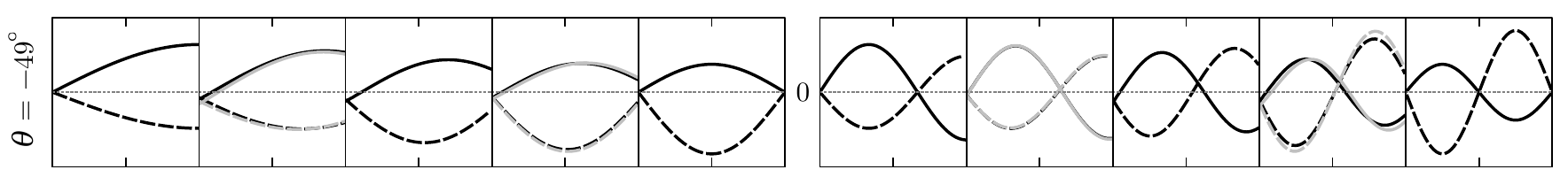}
  \includegraphics[scale=1.0]{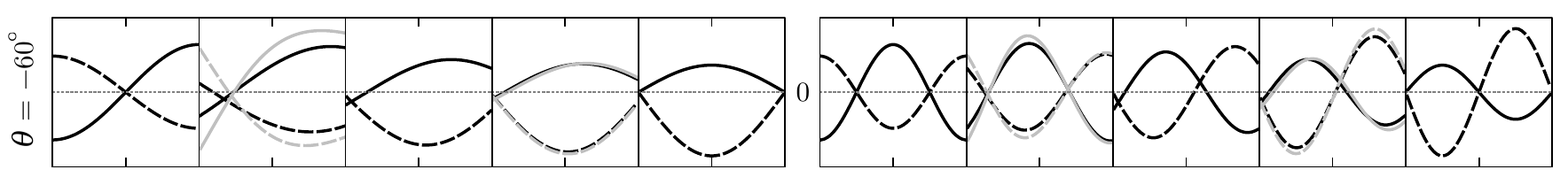}
  \includegraphics[scale=1.0]{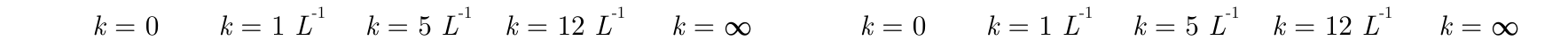}
  \caption{\label{fig:profiles} Profiles $\delta y(z)$ (solid lines) and $-i\delta z(z)$ (dashed lines), with $0 \le z \le L$, of the lowest-frequency volume modes $n=1,2$, for a range of wavevectors $\mathbf{k} = (k\cos\vartheta,k\sin\vartheta)$, taking $H_x = 0.73\,M_\text{S}$ and $2K = 0.46\,\mu_0 M_\text{S}^2$.
  The modes are invariant under a reflection of $\mathbf{k}$ with respect to the $k_y$ axis ($\vartheta \leftrightarrow 180^\circ - \vartheta$).
  Notice that the limiting profile for $k=0$ depends on the direction~$\vartheta$ from which we approach the singularity at $\mathbf{k}=0$ \cite{Hurben1995}.
  The critical angle $\vartheta_\text{cr}=49^\circ$ defines the boundary between regions~A (uniform limiting profile for $n=1$) and~B (sinusoidal limiting profile), as shown schematically in \figref{fig:regsketch}.
  For $k = 1\,L^{-1}$ and $k = 12\,L^{-1}$, we compare the numerical solutions (black lines) of the normal-mode problem~\eqref{eq:modez} to the first-order approximations (gray lines) given by \equarefs{eq:genform}, \eqref{eq:expandsmallk}, \eqref{eq:expandbigk}, and \tabref{tab:modeexpr}.
  Our first-order expressions provide a good indication of the numerical mode profiles, not only in the $k\rightarrow0$ or $k\rightarrow\infty$ limits \cite{Hurben1995} but also for finite $k$.
  }
\end{figure*}

\Figref{fig:profiles} shows the mode profiles $\delta y(z), \delta z(z)$ of the two lowest volume modes $n=1,2$ for a range of wavevectors~$\mathbf{k} = (k_x,k_y) = (k\cos\vartheta,k\sin\vartheta)$.
While we find that the profiles do not depend in any way on the sign of $k_x$, notice that the cases $k_y>0$ and $k_y<0$ are inequivalent \cite{Hurben1995}.
In particular, the antinode (amplitude maximum) of the $n=1$ mode tends to move towards one or the other surface of the film ($z=0$ or $z=L$) depending on the sign of $k_y$. A similar nonreciprocity is seen in the DE modes, which exponentially localize near either of the two film surfaces \cite{DamonEshbachPhys}.
The explicit perturbative expressions for the mode profiles, which we present in \secref{sec:limbeh}, can be used to quantify the asymmetric behavior.

\subsection{Relation to uniform-mode analysis}

It is interesting to compare the numerical dispersion relation of the $n=1$ volume mode to the dispersion relation~\eqref{eq:unifdisp} obtained in the uniform-mode approximation.
As shown in \figref{fig:dispavoid}(b), we find that \equaref{eq:unifdisp} predicts the correct group velocity $\ddnoskip{}\omega / \ddnoskip k$ in the $k\rightarrow0$ limit when approaching the point $\mathbf{k}=0$ along the $k_x$ axis ($\vartheta = 0^\circ$ or $\vartheta = 180^\circ$).
Along the $k_y$ axis ($\vartheta = \pm90^\circ$), however, the numerical dispersion relation differs very significantly from the uniform-mode expression. In particular, the slope $\ddnoskip{}\omega / \ddnoskip k$ predicted for $k\rightarrow0$ is incorrect: we have $\ddnoskip{}\omega / \ddnoskip k \rightarrow 0$ for the volume modes, but \equaref{eq:unifdisp} predicts a positive group velocity.
A qualitative explanation for the discrepancy may be found in \figref{fig:profiles}.
Approaching $\mathbf{k}=0$ along the $k_x$ axis ($\vartheta = 0^\circ$), it is found \cite{Hurben1995} that the limiting profile $\psi_0(z)$ of the $n = 1$ mode is indeed a constant function on the interval $0\le z \le L$, as was assumed in the uniform-mode approach.
Along the $k_y$ axis ($\vartheta = 90^\circ$), by contrast, we have that $\psi_0(z)$, for $0\le z \le L$, is a cosine function with wavenumber $\pi / L$, and as a result, the uniform-mode analysis is inaccurate even for small~$k$.
Regardless of~$\vartheta$, the uniform-mode analysis is also inaccurate in the large-$k$ regime (limiting profile for $k\rightarrow\infty$ is a sine function). However, the limiting frequency~\eqref{eq:freqinf} is reproduced correctly.

Along the $k_y$ axis, the uniform-mode dispersion relation~\eqref{eq:unifmode} coincides, in the small-$k$ regime, with the DE curve.
The DE mode, which is exponentially localized to the surface with a decay rate proportional to $k_y$ \cite{DamonEshbachPhys}, assumes a uniform profile in the limit $k\rightarrow0$.
In other words, the uniform profile, which corresponds to the lowest-frequency mode ($n=1$) for $\vartheta = 0^\circ$, becomes the highest-frequency mode (DE) for $\vartheta = 90^\circ$.
At the same time, the profile of the $n=1$ volume mode goes from uniform ($\vartheta = 0^\circ$) to sinusoidal ($\vartheta = 90^\circ$).
We might interpret the transition as an avoided band crossing, as shown in \figref{fig:dispavoid}(c).
The reason for the dependence of the limiting profile $\psi_0(z)$ on $\vartheta$ is given in more formal terms in \secref{sec:limbeh}.

\section{\label{sec:limbeh}Limiting behavior}

In this section, we present explicit analytical expressions for the frequency $\omega$ and profiles $\delta y(z), \delta z(z)$ of the volume modes $n$ in the small-$k$ and large-$k$ regimes.
While the limiting profiles for $k\rightarrow0$ and $k\rightarrow\infty$ (zeroth order) are well known \cite{Hurben1995}, our expressions, which are accurate up to first order in $k$ or $1/k$, give a good impression of the behavior of the modes even for finite $k$, as shown in \figref{fig:profiles}.
They can be used to estimate how strongly each volume mode $n$ couples to an external field pulse with a given depth profile, or to predict the contribution of the mode to net magnetization $\delta z(t,x,y) = \int \delta z (t,x,y,z) \dd{}z$ as measured using Faraday rotation \cite{Satoh2012}.
They also describe quantitatively the asymmetry in the profiles obtained for $k_y\neq0$.
Moreover, we use the perturbation theory derived here to construct an accurate semianalytical expression for the dispersion relation of the $n=1$ volume mode in \secref{sec:semianalytical}.

\begin{figure}
  \includegraphics[scale=1.0]{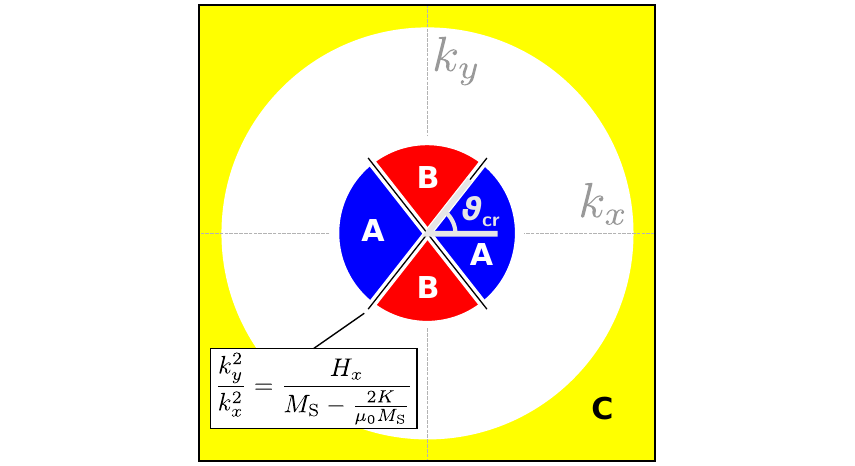}
  \caption{\label{fig:regsketch}(color online). Domains of applicability of the three asymptotic regimes A, B, and C.
  Regions A and B together represent the small-$k$ regime.
  The behavior of the normal modes and frequencies is qualitatively different depending on whether one approaches the point $\mathbf{k}=0$ from region A or region B.
  Region C denotes the large-$k$ regime.
  }
\end{figure}

The limiting behavior for $k\rightarrow0$ depends essentially on the polar angle $\vartheta$.
It is useful to introduce the quantity
\begin{equation}\label{eq:exprH}
H = H_x\cos^2\vartheta - \bigl(M_\text{S} - \tfrac{2K}{\mu_0 M_\text{S}}\bigr)\sin^2\vartheta
\text{.}
\end{equation}
The boundary lines $H=0$ separate the small-$k$ domain into four sectors, as shown in \figref{fig:regsketch}.
We distinguish between region~A, where $H>0$, and region~B, where $H<0$.
Regions~A and~B meet at the critical angle $\vartheta_\text{cr} = \arctan\sqrt{H_x / (M_\text{S} - \tfrac{2K}{\mu_0 M_\text{S}})}$ \cite{Hurben1995}.
The large-$k$ domain is designated as region~C.

It is convenient to write the mode profiles as
\begin{equation}\label{eq:genform}
\Psi_+
=
\left(
\begin{array}{c}
\delta y \\
\delta z 
\end{array}
\right)
=
\sqrt{\frac{\mu_0 |\gamma|}{2 ab}}
\left(
\begin{array}{c}
\phantom{-i}a \, [\psi(z) - \phi(z)] \\
-ib \, [\psi(z) + \phi(z)]
\end{array}
\right)
\text{,}
\end{equation}
where $\psi(z),\phi(z)$ are real-valued functions supported on the interval $0 \le z \le L$.
The constants $a,b>0$ are defined by \equaref{eq:absmallk} for small $k$ (regions~A and~B) and by \equaref{eq:abbigk} for large $k$ (region~C).
The normalization condition~\eqref{eq:normconstraint} becomes
\begin{equation}\label{eq:normpsiphi}
\Psi_+^\dagger \hat{Q} \Psi_+ 
= \int_{0}^{L}\bigl[\psi(z)^2 - \phi(z)^2\bigr]\dd{}z
= 1
\text{.}
\end{equation}
In the small-$k$ regime (regions~A and~B), we expand the wavefunctions and eigenfrequencies as
\begin{subequations}\label{eq:expandsmallk}
\begin{align}
\omega& = \omega_0 + k\omega_1 + k^2\omega_2 + \ldots\text{,}\\
\psi(z)& = \psi_0(z) + k\psi_1(z) + \Or(k^2) \text{,}\\
\phi(z)& = k\phi_1(z) + \Or(k^2)
\text{;}
\end{align}
\end{subequations}
in the large-$k$ regime (region C), we define
\begin{subequations}\label{eq:expandbigk}
\begin{align}
\omega& = \omega_0 + k^{-1}\omega_1 + k^{-2}\omega_2 + \ldots\text{,}\\
\psi(z)& = \psi_0(z) + k^{-1}\psi_1(z) + \Or(k^{-2}) \text{,}\\
\phi(z)& = k^{-1}\phi_1(z) + \Or(k^{-2})
\text{.}
\end{align}
\end{subequations}
In all three regions, only the $\psi(z)$ component of the wavefunction contributes at zeroth order ($k=0$ or $k=\infty$); the function $\phi(z)$ vanishes in those limits [see \equaref{eq:asympprof0}].

The main results of this section are summarized in \tabrefs{tab:asymp} and~\ref{tab:modeexpr}, 
which list explicit perturbative expressions for frequency (up to second order) and profiles (up to first order) of the volume modes $n$, for each of the regions.
For brevity, we introduce the quantities
\begin{subequations}\label{eq:exprAGJ}
\begin{align}
A& = \bigl(H_x + M_\text{S} - \tfrac{2K}{\mu_0 M_\text{S}}\bigr)\sin^2\vartheta + H_x \text{,} \\
G& = H_x \cos^2\vartheta + \tfrac{2K}{\mu_0 M_\text{S}}\sin^2\vartheta \text{,} \\
J& = H_x\cos^2\vartheta + (\tfrac{2K}{\mu_0 M_\text{S}} + M_\text{S})\sin^2\vartheta \text{.}
\end{align}
\end{subequations}
The asymptotic behavior of the dispersion relations, given by the expressions in \tabref{tab:asymp}, is shown for $n=1$ in \figref{fig:dispavoid}(b).
In \figref{fig:profiles}, we successfully compare our first-order mode profiles, given by the expressions in \tabref{tab:modeexpr}, to the numerical results.

In the remainder of this section, we present in more detail the derivations for each of the regions~A (\secref{sec:regA}), B (\secref{sec:regB}), and C (\secref{sec:regC}). The boundary between regions~A and~B, where $|\vartheta| = \vartheta_\text{cr}$ or $|\vartheta| = 180^\circ - \vartheta_\text{cr}$, requires special consideration (\secref{sec:regAB}).
In the interest of readability, we focus on the limiting profiles $\psi_0(z)$ of the $n=1$ mode and on the asymptotic behavior of its dispersion relation.
A more mathematical derivation of the perturbation theory used to obtain all results in \tabrefs{tab:asymp} and~\ref{tab:modeexpr} is given in Appendix~\ref{sec:pertthy}.

\begin{table*}
\caption{\label{tab:asymp}Asymptotic behavior of the dispersion relations of the volume modes $n$, in each of the long-wavelength regions~A, B, and the A--B boundary line and in the short-wavelength region C (see \figref{fig:regsketch}). The $\vartheta$-dependent quantities $H$, $J$, and $G$ are defined in \equarefs{eq:exprH} and~\eqref{eq:exprAGJ}. The profiles $\psi_0,\psi_1,\phi_1$ (see \tabref{tab:modeexpr}) are shown for $n=1$ (black lines) and $n=2$ (gray lines).
}
\begin{tabular}{p{0.22\linewidth} | p{0.06\linewidth} | p{0.37\linewidth} | c | c c}
 & & $\dfrac{\omega}{|\gamma|\mu_0} = ab + \ldots $ & $\psi_0(z)$  & $\psi_1(z)$ & $\phi_1(z)$ \\[0.5em] \hline
 \multirow{4}{*}{$\left.\begin{array}{l} \\ ~\\ ~ \\ a= \sqrt{H_x + M_\text{S} -\tfrac{2K}{\mu_0 M_\text{S}}} \\ b= \sqrt{H_x} \\ ~ \\ ~ \end{array}\right\lbrace$} & \centering \textbf{A} ($n=1$) & $-  \dfrac{M_\text{S} H}{4ab} kL
+ \dfrac{M_\text{S}}{12ab} \Bigl(  J  - \dfrac{3 M_\text{S} H^2}{8 a^2 b^2} \Bigr) (kL)^2 \newline \hspace*{15em}
+ \mathcal{O}(k^3/H)$ 
 & \parbox{0.10\linewidth}{\includegraphics[scale=1.0]{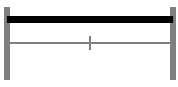}} 
 & \parbox{0.10\linewidth}{\includegraphics[scale=1.0]{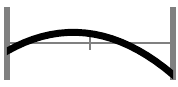}}
 & \parbox{0.10\linewidth}{\includegraphics[scale=1.0]{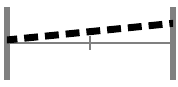}} 
 \\[0.5em] 
   & \centering \textbf{A} ($n>1$) & $- \Bigl(\dfrac{kL}{(n-1)\pi}\Bigr)^2 \dfrac{M_\text{S} G}{2 a b } + \mathcal{O}(k^3/H)$ 
 & \parbox{0.10\linewidth}{\includegraphics[scale=1.0]{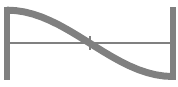}} 
 & \parbox{0.10\linewidth}{\includegraphics[scale=1.0]{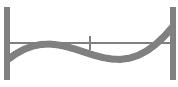}}
 & \parbox{0.10\linewidth}{\includegraphics[scale=1.0]{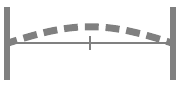}} 
 \\[0.5em] 
  & \centering \textbf{A}--\textbf{B} & $  
  -\Bigl(\dfrac{2kL}{(2n-1)\pi}\Bigr)^2
  \dfrac{ M_\text{S} G_0}{2 ab } q_n(\eta) + \mathcal{O}(|\eta|k^3 + k^3) $
 & \parbox{0.10\linewidth}{\includegraphics[scale=1.0]{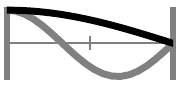}} 
 & \parbox{0.10\linewidth}{\includegraphics[scale=1.0]{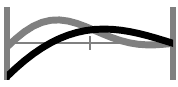}}
 & \parbox{0.10\linewidth}{\includegraphics[scale=1.0]{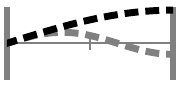}} 
  \\[0.5em] 
  & \centering \textbf{B} & $ - \Bigl(\dfrac{kL}{n\pi}\Bigr)^2 \dfrac{M_\text{S} G}{2 a b }
+ \mathcal{O}[k^3/(-H)]$
 & \parbox{0.10\linewidth}{\includegraphics[scale=1.0]{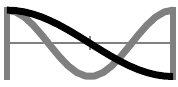}} 
 & \parbox{0.10\linewidth}{\includegraphics[scale=1.0]{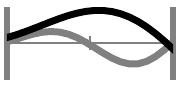}}
 & \parbox{0.10\linewidth}{\includegraphics[scale=1.0]{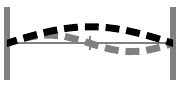}} 
  \\[0.5em] 
   $\begin{array}{l} a= \sqrt{H_x - \frac{2K}{\mu_0 M_\text{S}}} \\ b= \sqrt{H_x + M_\text{S}\sin^2 \vartheta} \end{array}$ & \centering \textbf{C} & $ +\Bigl(\dfrac{n \pi}{kL}\Bigr)^2 \dfrac{ M_\text{S} G}{2ab} + \Or(k^{-3}) $
 & \parbox{0.10\linewidth}{\includegraphics[scale=1.0]{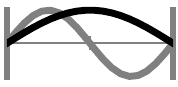}} 
 & \parbox{0.10\linewidth}{\includegraphics[scale=1.0]{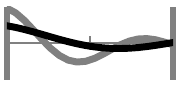}}
 & \parbox{0.10\linewidth}{\includegraphics[scale=1.0]{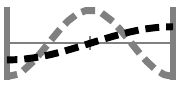}} 
\end{tabular}
\end{table*}

\begin{table}
\caption{\label{tab:modeexpr}Mode profiles and first-order corrections.
All functions to be multiplied by $\Pi^*(z/L)$. See \equarefs{eq:genform}, \eqref{eq:expandsmallk}, \eqref{eq:expandbigk}.
}
\phantom{a}
\begin{minipage}[t]{\linewidth}
\begin{flushleft}
\textbf{Region A ($n=1$)}
\end{flushleft}
\end{minipage}
\begin{tabular}{@{}c@{} p{0.80\linewidth}}
  \parbox{57pt}{\includegraphics[scale=1.0]{A1psi0}} 
  & $\begin{aligned}\psi_0(z) &= \dfrac{1}{\sqrt{L}}\end{aligned}$ \\
  \parbox[t]{57pt}{\includegraphics[scale=1.0]{A1psi1}} 
  & $\begin{aligned}\psi_1(z) &= -\dfrac{1}{H\sqrt{L}} \\
  & \phantom{={}}\!\!\!\!\!\!\!\!\!\!\!\!\times \Bigl[   \dfrac{AM_\text{S} \sin\vartheta}{2ab}  \dfrac{2z-L}{2} +  G  \dfrac{3(2z-L)^2 - L^2}{12L} \Bigr] \end{aligned}$ \\
  \parbox{57pt}{\includegraphics[scale=1.0]{A1phi1}} 
  & $\begin{aligned}\phi_1(z) &=  \dfrac{1}{\sqrt{L}} \dfrac{M_\text{S}}{2ab} \Bigl(  \dfrac{AL}{4ab} + \sin\vartheta \dfrac{2z - L}{2}  \Bigr) \end{aligned}$\\
\end{tabular}
\begin{minipage}[t]{\linewidth}
\phantom{a}
\begin{flushleft}
\textbf{Region A ($n>1$)}
\end{flushleft}
\end{minipage}
\begin{tabular}{@{}c@{} p{0.80\linewidth} }
  \parbox{57pt}{\includegraphics[scale=1.0]{Anpsi0}} 
  & $\begin{aligned}\psi_0(z) &= \sqrt{\dfrac{2}{L}} \cos\Bigl(\dfrac{(n-1)\pi z}{L}\Bigr) \end{aligned}$ \\
  \raisebox{\height}{\parbox[t]{57pt}{\includegraphics[scale=1.0]{Anpsi1}}} 
  & $\begin{aligned}\psi_1(z)& = - \sqrt{\dfrac{2}{L}} \dfrac{1}{H} \dfrac{L}{(n-1)\pi}  \\
 & \phantom{={}} \times \biggl\langle \dfrac{AM_\text{S} \sin\vartheta}{2ab}   \sin\Bigl(\dfrac{(n-1) \pi z}{L}\Bigr) \\
 & \phantom{=\times \biggl\langle}+ G \Bigl[ \dfrac{2z-L}{L} \sin\Bigl(\dfrac{(n-1)\pi z}{L}\Bigr) \\
 & \phantom{=\times \biggl\langle+ G \Bigl[}+ \dfrac{1}{(n-1) \pi}\cos\Bigl(\dfrac{(n-1) \pi z}{L}\Bigr)\Bigr] \biggr\rangle
 \end{aligned}$ \\
  \parbox{57pt}{\includegraphics[scale=1.0]{Anphi1}} 
  & $\begin{aligned}\phi_1(z) &=  \sqrt{\dfrac{2}{L}}  \dfrac{M_\text{S}\sin\vartheta}{2ab}  \dfrac{L}{(n-1)\pi} \sin\Bigl(\dfrac{(n-1)\pi z}{L}\Bigr)\end{aligned}$\\
\end{tabular}
\begin{minipage}[t]{\linewidth}
\phantom{a}
\begin{flushleft}
\textbf{Boundary A--B (for $\eta=0$ and $k_y>0$)}\textsuperscript{a}
\end{flushleft}
\end{minipage}
\begin{tabular}{@{}c@{} p{0.80\linewidth} }
  \parbox{57pt}{\includegraphics[scale=1.0]{ABpsi0}} 
  & $\begin{aligned}\psi_0(z) &= \sqrt{\dfrac{2}{L}} \cos\Bigl(\dfrac{(2n-1) \pi z}{2L}\Bigr)\end{aligned}$ \\
  \parbox[t]{57pt}{\includegraphics[scale=1.0]{ABpsi1}} 
  & $\begin{aligned}
  \psi_1(z)& =- \sqrt{\dfrac{2}{L}}   \dfrac{2L}{(2n-1)\pi} \Bigl[ \dfrac{z-L}{L} \sin\Bigl(\dfrac{(2n-1) \pi z}{2L}\Bigr) \\
  &\phantom{={}}\quad\quad + \dfrac{1}{(2n-1)\pi} \cos\Bigl(\dfrac{(2n-1) \pi z}{2L}\Bigr) 
\Bigr]{}& 
  \end{aligned}$ \\
  \parbox{57pt}{\includegraphics[scale=1.0]{ABphi1}} 
  & $\begin{aligned}\phi_1(z) &= \sqrt{\dfrac{2}{L}} \dfrac{M_\text{S}}{2a^2} \dfrac{2L}{(2n-1) \pi }  \sin\Bigl(\dfrac{(2n-1) \pi z}{2L}\Bigr)\end{aligned}$
\end{tabular}
\begin{minipage}[t]{\linewidth}
\begin{flushleft}
\textsuperscript{a}~For $k_y<0$, take $\psi_0(z) \leftarrow (-1)^{n-1}\psi_0(L-z)$ for $\psi_0,\psi_1, \phi_1$.
\end{flushleft}
\end{minipage}
\begin{minipage}[t]{\linewidth}
\phantom{a}
\begin{flushleft}
\textbf{Region B}\textsuperscript{b}
\end{flushleft}
\end{minipage}
\begin{tabular}{@{}c@{} p{0.80\linewidth} }
  \parbox{57pt}{\includegraphics[scale=1.0]{Bpsi0}} 
  & $\begin{aligned}\psi_0(z) &= \pm \sqrt{\dfrac{2}{L}} \cos\Bigl(\dfrac{n \pi z}{L}\Bigr) \end{aligned}$ \\
  \parbox[t]{57pt}{\includegraphics[scale=1.0]{Bpsi1}} 
  &$\begin{aligned}\psi_1(z)& = \mp \sqrt{\dfrac{2}{L}} \dfrac{1}{H} \dfrac{L}{n\pi} \biggl\langle \dfrac{AM_\text{S} \sin\vartheta}{2ab}   \sin\Bigl(\dfrac{n \pi z}{L}\Bigr) \\
 & \phantom{={}}\!\!\!\!\!\!\!\!\!\!+ G \Bigl[\dfrac{2z-L}{L} \sin\Bigl(\dfrac{n\pi z}{L}\Bigr) + \dfrac{1}{n\pi}\cos\Bigl(\dfrac{n \pi z}{L}\Bigr) \Bigr] \biggr\rangle
 \end{aligned}
 $ \\
  \parbox{57pt}{\includegraphics[scale=1.0]{Bphi1}} 
  & $\begin{aligned}\phi_1(z) &= \sqrt{\dfrac{2}{L}} \dfrac{M_\text{S} |\sin\vartheta|}{2 a b} \dfrac{L}{n \pi} \sin\Bigl(\dfrac{n \pi z}{L}\Bigr)\end{aligned}$
\end{tabular}
\begin{minipage}[t]{\linewidth}
\begin{flushleft}
\textsuperscript{b}~For $k_y>0$ (upper signs) and $k_y<0$ (lower signs).
\end{flushleft}
\end{minipage}
\begin{minipage}[t]{\linewidth}
\phantom{a}
\begin{flushleft}
\textbf{Region C}
\end{flushleft}
\end{minipage}
\begin{tabular}{@{}c@{} p{0.80\linewidth} }
  \parbox{57pt}{\includegraphics[scale=1.0]{Cpsi0}} 
  & $\begin{aligned}\psi_0(z) &= \sqrt{\dfrac{2}{L}} \sin\Bigl(\dfrac{n \pi z}{L}\Bigr)\end{aligned}$ \\
  \parbox[t]{57pt}{\includegraphics[scale=1.0]{Cpsi1}} 
  & $\begin{aligned}
  \psi_1(z) &=  \sqrt{\dfrac{2}{L}} \dfrac{1}{J} \dfrac{n \pi}{L} \biggl\langle 
  \dfrac{AM_\text{S} \sin\vartheta}{2ab} \cos\Bigl(\dfrac{n \pi z}{L}\Bigr) \\
  &\phantom{={}}\!\!\!\!\!\!\!\!\!\! -  G  \Bigl[
  \dfrac{2z - L}{L} \cos\Bigl(\dfrac{n \pi z}{L}\Bigr)
  +  \dfrac{1}{n \pi}  \sin\Bigl(\dfrac{n \pi z}{L}\Bigr)
\Bigr]
\biggr\rangle
  \end{aligned}$ \\
  \parbox{57pt}{\includegraphics[scale=1.0]{Cphi1}} 
  & $\begin{aligned}\phi_1(z) &= - \sqrt{\dfrac{2}{L}} \dfrac{M_\text{S} \sin\vartheta}{2ab} \dfrac{n \pi}{L} \cos\Bigl(\dfrac{n \pi z}{L}\Bigr)\end{aligned}$
\end{tabular}
\end{table}

\subsection{\label{sec:regA}Region A}

The operators $\hat{D}^{ab}(k_x,k_y)$, defined in \equaref{eq:defD}, are the only nontrivial operators appearing in the eigenvalue equation~\eqref{eq:modez}.
We expand $\hat{D}^{ab}(k\cos\vartheta,k\sin\vartheta)$ in the parameter $k$ (for fixed $\vartheta$).
Using Appendix~\ref{sec:distlimit}, we obtain
\begin{multline}\label{eq:Dexpsmall}
\left(
\begin{array}{c c}
\hat{D}^{yy} & \hat{D}^{yz} \\
\hat{D}^{yz} & \hat{D}^{zz} \\
\end{array}
\right)
=
\hat{D}_0 + k\hat{D}_1 + k^2\hat{D}_2 + k^3\hat{D}_3 + \ldots
\\=
\left(
\begin{array}{c c}
0  & 0 \\
0 & \hat{S} \\
\end{array}
\right)
+
k
\left(
\begin{array}{c c}
\pi (\sin^2\vartheta) \delta(\hat{k}_z)  & (\sin\vartheta) \hat{k}_z^{-1} \\
(\sin\vartheta) \hat{k}_z^{-1} & - \pi \delta(\hat{k}_z) \\
\end{array}
\right) 
\\+
k^2
\left(
\begin{array}{c c}
 (\sin^2\vartheta)\hat{k}_z^{-2}  & \pi (\sin\vartheta) \delta'(\hat{k}_z) \\
 \pi (\sin\vartheta) \delta'(\hat{k}_z) & -\hat{k}_z^{-2} \\
\end{array}
\right) 
\\+
k^3
\left(
\begin{array}{c c}
 -\tfrac{1}{2}\pi(\sin^2\vartheta) \delta''(\hat{k}_z)  &  -(\sin\vartheta)\hat{k}_z^{-3}  \\
 -(\sin\vartheta)\hat{k}_z^{-3}  & \tfrac{1}{2}\pi \delta''(\hat{k}_z)  \\
\end{array}
\right) 
+ \Or(k^4)\text{,}
\end{multline}
where $\delta$ represents the Dirac delta distribution.
The expressions containing $\hat{k}_z$ represent (in real space) convolution operators acting on the profiles $\delta y(z), \delta y(z)$; for example, the action of $\hat{D}_1$ may be expressed as
\begin{multline}
\hat{D}_1
\left(
\begin{array}{c}
\delta y(z) \\
\delta z(z) \\
\end{array}
\right)
=  \frac{1}{2} \times \\
\left(
\begin{array}{c}
 \int [ \sin^2\vartheta \, \delta y(z')  + i \sin\vartheta \sign(z-z') \, \delta z(z') ] \dd{}z' \\
 \int [ i \sin\vartheta \sign(z-z') \, \delta y(z')  -  \delta z(z') ] \dd{}z' \\
\end{array}
\right)
\text{,}
\end{multline}
where we have used the Fourier transforms $f(z)=1 \leftrightarrow \tilde{f}(k_z)=2\pi\delta(k_z)$ and $f(z)=\sign(z) \leftrightarrow \tilde{f}(k_z)=-2i/k_z$.

All modes $n$ are degenerate for $k=0$, with $\omega_0 = |\gamma| \mu_0 ab$. The first- and second-order terms of the expansion of $\hat{D}^{ab}$ in $k$ lift this degeneracy and fix the spatial profile $\psi_0(z)$ in \equaref{eq:asympprof0}.
We obtain the profile of the lowest mode $n=1$ by minimizing
\begin{multline}\label{eq:smallkfunctional}
\omega_1 = \Psi_0^\dagger (M_\text{S}\hat{D}_1) \Psi_0 = \\
-\frac{|\gamma| \mu_0 M_\text{S}}{4ab}  \bigl[H_x\cos^2\vartheta - (M_\text{S} - \tfrac{2K}{\mu_0 M_\text{S}})\sin^2\vartheta\bigr] \\
\times
\biggl(\int_{0}^{L}\psi_0(z)\dd{}z\biggr)^2
\end{multline}
under the constraint $\int_{0}^{L}\psi_0(z)^2\dd{}z = 1$ [\equaref{eq:constr0neat}].

We identify the prefactor between square brackets in \equaref{eq:smallkfunctional} as the quantity $H$.
In region~A ($H>0$), minimization of $\omega_1$ is equivalent to maximization of $\bigl(\int_{0}^{L}\psi_0(z)\dd{}z\bigr)^2$, yielding the uniform profile
\begin{equation}
\psi_0(z) =
\frac{1}{\sqrt{L}}
\Pi^*\Bigl(\frac{z}{L}\Bigr)\text{.}
\end{equation}
By definition, $\omega_1 = \lim_{k\rightarrow0}\ddnoskip{} \omega / \ddnoskip{}k$ is the group velocity for $k=0$. We obtain
\begin{equation}
\frac{\ddnoskip{}\omega}{\ddnoskip{}k}
=
-\frac{\mu_0 |\gamma| M_\text{S} L}{4ab} H
+ \mathcal{O}(k)
\text{,}
\end{equation}
as in the uniform-mode approach [see \figref{fig:dispavoid}(b)].

Now that the $k=0$ profile $\psi_0(z)$ of the $n=1$ mode is known, 
the part $\phi_1(z)$ of the first-order correction is given by \equaref{eq:phi1sol}, which condition results from minimization of $\omega_2$.
However, the other part $\psi_1(z)$ does not affect the value of $\omega_2$, given by \equaref{eq:func2neat}, if the constraint $\int_{0}^{L} \psi_0(z) \psi_1(z) \dd{}z = 0$ is satisfied, and hence $\psi_1(z)$ cannot be determined by minimization of $\omega_2$.
This indeterminacy is a consequence of the degeneracy of the $\psi(z)$ component of the modes at zeroth order of perturbation theory.
We turn to minimization of $\omega_3$ to fix $\psi_1(z)$, yielding the condition~\eqref{eq:psi1sol}.
The resulting profiles are listed in \tabref{tab:modeexpr}.

\subsection{\label{sec:regB}Region B}

In region~B, where $H<0$, minimization of \equaref{eq:smallkfunctional} is equivalent to minimization of $\bigl(\int_{0}^{L}\psi_0(z)\dd{}z\bigr)^2$. The minimum value 
\begin{equation}
\omega_1 = \lim_{k\rightarrow0}\frac{\ddnoskip{}\omega}{\ddnoskip{}k}
=
0
\end{equation}
is obtained for any profile $\psi_0(z)$ for which $\int_{0}^{L}\psi_0(z)\dd{}z = 0$.
In other words, the zeroth-order profile $\psi_0(z)$ is indeterminate even in first-order perturbation theory. The degeneracy is lifted by the second-order term of \equaref{eq:Dexpsmall}.
By \equaref{eq:phi0func2nd}, the profile $\psi_0(z)$ minimizes
\begin{multline}\label{eq:func0regB}
\omega_2 = 
-\frac{|\gamma| \mu_0 M_\text{S}}{2ab} \bigl[ H_x\cos^2\vartheta +  \tfrac{2K}{\mu_0 M_\text{S}}\sin^2\vartheta  \bigr] \\
\times \int_{0}^{L} \psi_0 \hat{k}_z^{-2} \psi_0 \dd{}z
\text{.}
\end{multline}
The operator $\hat{k}_z^{-2}$ represents, as usual, a convolution in real space. Using the Fourier transform $\hat{f}(k)=k^{-n} \leftrightarrow f(x)=i(ix)^{n-1} \sign(x)/[2 (n-1)!]$, we have
\begin{multline}
\int_{0}^{L} \psi_0 \hat{k}_z^{-2} \psi_0 \dd{}z = \\ 
-\frac{1}{2}\int_{0}^{L} \psi_0(z) \int_{0}^{L} |z-z'|  \psi(z') \dd{}z' \dd{}z
\text{.}
\end{multline}
Minimization of $\omega_2$ gives
\begin{equation}
\psi_0(z) = 
\sqrt{\frac{2}{L}}
\cos\Bigl(\frac{\pi z}{L}\Bigr)\Pi^*\Bigl(\frac{z}{L}\Bigr)
\text{,}
\end{equation}
and we evaluate
\begin{multline}
\omega_2 = \lim_{k\rightarrow0} \frac{\ddnoskip{}\omega}{\ddnoskip{}(k^{2})}
=
\lim_{k\rightarrow0} \frac{1}{2k} \frac{\ddnoskip{}\omega}{\ddnoskip{}k}
\\=
-\frac{\mu_0|\gamma| M_\text{S} L^2}{2\pi^2 ab}  \bigl[H_x\cos^2\vartheta + \tfrac{2K}{\mu_0 M_\text{S}}\sin^2\vartheta \bigr]
\text{.}
\end{multline}
As in region~A, minimization of $\omega_2$ immediately fixes $\phi_1(z)$ according to \equaref{eq:phi1sol}.
However, the third-order expression~\eqref{eq:psi1sol} for $\psi_1(z)$ is indeterminate in region~B. We need to minimize the fourth-order functional $\omega_4$, given by \equaref{eq:om4func}, to determine $\psi_1(z)$.
The resulting expressions are listed in Table~\ref{tab:modeexpr}.

\subsection{\label{sec:regAB}Boundary line A--B}

The boundary line between regions A and B requires special consideration, as neither the perturbation theory of region A nor of region B is valid on this line.
We find that the the boundary line in some sense interpolates between the profiles in the interior of regions~A and~B.

On the boundary, where $H=0$, we have that $\omega_1$ as given by \equaref{eq:smallkfunctional} is identically zero.
This means that, as in region~B, the profile $\psi_0(z)$ is fixed by minimization of $\omega_2$. In contrast to region~B, however, there is no constraint $\int_{0}^{L}\psi_0(z)\dd{}z = 0$ from minimization of $\omega_1$.
By \equaref{eq:phi0func2nd}, $\psi_0(z)$ minimizes
\begin{equation}\label{eq:funcbound}
\omega_2 = - \frac{1}{2\omega_0} \int_{0}^{L} |(\hat{Y}_1^\dagger \psi_0)|^2 \dd{}z
\text{,}
\end{equation}
where
\begin{subequations}
\begin{align}
(\hat{Y}_1^\dagger \psi_0)(z)& = 
-C \int_{0}^{z} \psi_0(z') \dd{}z'
&\text{for $k_y>0$,}\\
(\hat{Y}_1^\dagger \psi_0)(z)& = 
C \int_{z}^{L} \psi_0(z') \dd{}z'
&\text{for $k_y<0$,}
\end{align}
\end{subequations}
with $C = \mu_0 |\gamma| M_\text{S} \sin\vartheta$.
Minimization gives
\begin{subequations}\label{eq:psi0AB}
\begin{align}
\psi_0(z)& = 
\sqrt{\frac{2}{L}} \cos\Bigl(\frac{\pi z}{2L} \Bigr) \Pi^*\Bigl(\dfrac{z}{L}\Bigr)
&\text{for $k_y>0$,}\\
\psi_0(z)& =  
\sqrt{\frac{2}{L}} \sin\Bigl(\frac{\pi z}{2L}\Bigr) \Pi^*\Bigl(\dfrac{z}{L}\Bigr)
&\text{for $k_y<0$}
\text{,}
\end{align}
\end{subequations}
and we evaluate
\begin{equation}\label{eq:om2AB}
\omega_2 = 
\lim_{k\rightarrow0} \frac{1}{2} \Bigl( \frac{\partial^2\omega}{\partial k^2} \Bigr)_{H=0}
=
-|\gamma|\mu_0 \dfrac{4L^2}{\pi^2} \dfrac{ M_\text{S} G_0}{2 ab }
\text{,}
\end{equation}
where
\begin{equation}
G_0 = G|_{\vartheta = \vartheta_\text{cr}} = 
\frac{b^2 M_\text{S} }{ a^2 }
\text{.}
\end{equation}
Minimization of $\omega_2$ also fixes $\phi_1(z)$ by \equaref{eq:phi1sol}.
The other first-order component $\psi_1(z)$ is again determined only at fourth order of perturbation theory.

\begin{figure}
  \includegraphics[scale=1.0]{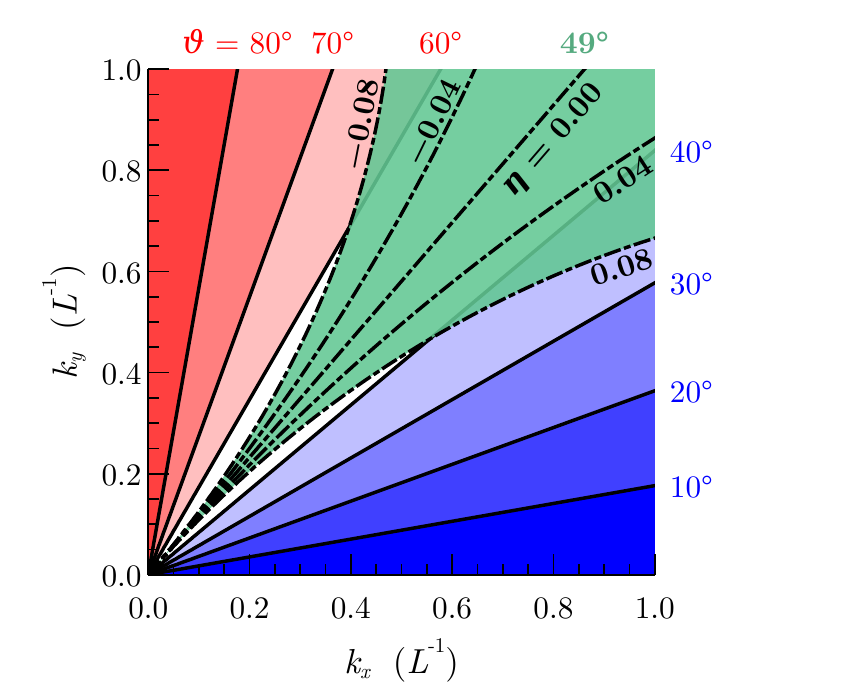}
  \caption{\label{fig:etatheta}(color online).
  In regions~A (blue) and~B (red), we carry out the expansion in $k$ along lines of constant $\vartheta$. In the A--B boundary region (green), we expand instead along curves of constant~$\eta$, as defined by \equaref{eq:defeta}. Together, the three perturbative expressions for $\omega$ (see \tabref{tab:asymp}) provide, up to the uncertainty in $q_n(\eta)$, a description of the dispersion relation that is accurate to second order in~$k$ uniformly in~$\vartheta$.
  }
\end{figure}

In the above expressions, we assume that we approach the point $\mathbf{k}=0$ along a line $H=0$; in other words, we fix $\vartheta = \pm \vartheta_\text{cr}$.
We can generalize \equaref{eq:om2AB} by carrying out the expansion along a curve of constant $\eta$, as shown in \figref{fig:etatheta}, where we define
\begin{equation}\label{eq:defeta}
\eta = \biggl(\frac{a^2}{2 b^2 M_\text{S}L} \biggr) \frac{H}{ k}
\text{.}
\end{equation}
We obtain, for the lowest mode $n=1$,
\begin{equation}\label{eq:om2abeta}
\omega_2 = \lim_{k\rightarrow0} \frac{1}{2} \Bigl( \frac{\partial^2 \omega}{\partial k^2} \Bigr)_\eta
=
-|\gamma|\mu_0 \dfrac{4L^2}{\pi^2} \dfrac{ M_\text{S} G_0}{2 ab } q_1(\eta)
\text{.}
\end{equation}
Notice that \equaref{eq:om2AB} corresponds to $\eta = 0$.
The generalization $\eta \neq 0$ interpolates between regions~A ($\eta\rightarrow\infty$) and~B ($\eta\rightarrow-\infty$)
and allows one, in principle, to construct a second-order approximation of $\omega$ around $k=0$ uniform in $\vartheta$.

\Tabref{tab:asymp} gives $\omega_2$ for arbitrary mode index $n$.
While we are unaware of a closed-form expression for the functions $q_n(\eta)$,
we have a small-$k$ expansion
\begin{equation}
q_n(\eta) = 1 + 2\eta + \eta^2 - \tfrac{1}{6}(2n-1)^2\pi^2 \eta^3(1-\eta) + \mathcal{O}(\eta^5)
\end{equation}
and large-$k$ expansions
\begin{subequations}
\begin{align}
q_n(\eta) &= \frac{\pi^2}{12}(1+3\eta) + \mathcal{O}(\eta^{-1}) & \text{for $\eta>0,n=1$,} \\
q_n(\eta) &= \Bigl(\frac{n-\tfrac{1}{2}}{n-1}\Bigr)^2 + \mathcal{O}(\eta^{-1}) & \text{for $\eta>0,n>1$,} \\
q_n(\eta) &= \Bigl(\frac{n-\tfrac{1}{2}}{n}\Bigr)^2 + \mathcal{O}(-\eta^{-1}) & \text{for $\eta<0$.}
\end{align}
\end{subequations}
A very good approximation for $q_1(\eta)$, with a maximal absolute error of $0.00363$, is given by
\begin{multline}
q_1(\eta) \approx
\frac{1}{24}  \Bigl[
3 + \pi^2(1+3\eta) \\
+ \sqrt{[3 + \pi^2(1+3\eta)]^2 - 12 \pi^2(1+3\eta+3)  + 432}
\Bigr]
\text{.}
\end{multline}

\subsection{\label{sec:regC}Region C}

For large $k$, we have the expansion [\emph{cf.} \equaref{eq:Dexpsmall}]
\begin{multline}\label{eq:expinf}
\left(
\begin{array}{c c}
\hat{D}_{yy} & \hat{D}_{yz} \\
\hat{D}_{zy} & \hat{D}_{zz} \\
\end{array}
\right)
=
\hat{D}_{-0} + k^{-1}\hat{D}_{-1} + k^{-2}\hat{D}_{-2} + \ldots \\
=
\sin^2\vartheta
\left(
\begin{array}{c c}
\hat{S} & 0 \\
0 & 0 \\
\end{array}
\right)
+ 
\frac{\sin\vartheta}{k}
\left(
\begin{array}{c c}
0 & \hat{k}_z \\
\hat{k}_z & 0 \\
\end{array}
\right)\\
+ \frac{1}{k^2}
\left(
\begin{array}{c c}
-(\sin^2\vartheta) \hat{k}_z^2 & 0 \\
0 & \hat{k}_z^2 \\
\end{array}
\right)
 - 
\frac{\sin\vartheta}{k^3}
\left(
\begin{array}{c c}
0 & \hat{k}_z^3 \\
\hat{k}_z^3 & 0 \\
\end{array}
\right)
\\
- \frac{1}{k^4}
\left(
\begin{array}{c c}
-(\sin^2\vartheta) \hat{k}_z^4 & 0 \\
0 & \hat{k}_z^4 \\
\end{array}
\right)
+\Or(k^{-5})
\text{.}
\end{multline}
All modes $n$ are degenerate at zeroth order.
In the $k \rightarrow \infty$ limit, the mode profiles are of the form~\eqref{eq:asympprof0} with $a,b$ given by \equaref{eq:abbigk}. Notice that, for region~C, the value of $b$ depends on $\vartheta$.
The second-order term $\hat{D}_{-2}$ in \equaref{eq:expinf} lifts the degeneracy and fixes the spatial profile $\psi_0(z)$.

Regardless of $k_y$, we have 
\begin{equation}
\lim_{k\rightarrow\infty} -k^2 \frac{\ddnoskip{} \omega }{ \ddnoskip{}k}
= \omega_1 = \Psi_0^\dagger (M_\text{S} \hat{D}_{-1}) \Psi_0 
= 0
\text{.}
\end{equation}
For $k_y=0$, the first-order term $\hat{D}_{-1}$ in \equaref{eq:expinf} even vanishes identically.
In this case, we may somewhat simplify our calculations by treating $\omega_2$ as the first-order term of a perturbation series in $k^2$.
We find the limiting profile $\psi_0(z)$ of the $n=1$ mode by minimization of 
\begin{equation}\label{eq:funcinfth0}
\omega_2 = \Psi_0^\dagger (M_\text{S}\hat{D}_{-2}) \Psi_0
=
\frac{\mu_0 |\gamma| M_\text{S}}{2ab}
H_x
\int_{0}^{L}\left(\frac{\ddnoskip{}\psi_0}{\ddnoskip{}z}\right)^2\dd{}z
\text{,}
\end{equation}
under the constraint~\eqref{eq:constr0neat}.
We obtain
\begin{equation}\label{eq:psi0kinfprofile}
\psi_0(z) = 
\sqrt{\frac{2}{L}}
\sin\Bigl(\frac{\pi z}{L}\Bigr) \Pi^*\Bigl(\frac{z}{L}\Bigr)
\end{equation}
and
\begin{equation}
\lim_{k\rightarrow\infty} -\frac{k^3}{2} \frac{\ddnoskip{} \omega }{ \ddnoskip{}k} =
\omega_2 = \mu_0 |\gamma| M_\text{S} \frac{\pi^2}{L^2} \frac{H_x}{2ab}
\text{.}
\end{equation}
Notice that $\psi_0(z)$ satisfies Dirichlet boundary conditions $\psi_0(0) = \psi_0(L) = 0$.
Such conditions are necessary to give \equaref{eq:funcinfth0} a finite value, since $\psi_0(z)$ must vanish outside the interval $0\le z \le L$.
However, higher-order terms $\phi_1(z),\psi_1(z)$ of the expansion can have finite values for $z=0$ or $z=L$.

In the general case $k_y\neq0$, the profile $\psi_0(z)$ is still a sine function~\eqref{eq:psi0kinfprofile}.
Using \equaref{eq:phi0func2nd}, we evaluate
\begin{equation}\label{eq:psi0kinf}
\omega_2
= 
\frac{\mu_0 |\gamma| M_\text{S} G}{2ab}
\int_{0}^{L}\Bigl(\frac{\ddnoskip{}\psi}{\ddnoskip{}z}\Bigr)^2\dd{}z 
 = 
\mu_0 |\gamma| M_\text{S} \frac{\pi^2}{L^2}
\frac{G}{2ab}
\text{.}
\end{equation}
The profile $\phi_1(z)$ is fixed by \equaref{eq:phi1sol}.
The other first-order profile $\psi_1(z)$ is determined, again, only by minimization of the fourth-order functional $\omega_4$ [\equaref{eq:om4func}].
\Tabref{tab:modeexpr} lists the resulting expressions.

When performing the derivation of the profile $\psi_1(z)$, we take into account the following. Writing out \equarefs{eq:om3func} and~\eqref{eq:om4func} for region~C, we find that the functionals $\omega_3$ and $\omega_4$ contain terms such as $i\int \psi_1 \hat{k}_z \phi_1 \dd{}z = \int \psi_1 \partial_z \phi_1 \dd{}z$ or $\int \psi_0 \hat{k}_z^4 \psi_0 \dd{}z = \int \psi_0 \partial_z^4 \psi_0 \dd{}z$, which must be regularized; indeed, the profiles $\phi_0(z), \psi_1(z)$ have discontinuities at $z=0$ and $z=L$, while $\psi_0(z)$ has discontinuities in its first derivative.
The functionals $\omega_3, \omega_4$ can each be written as a sum of regular integral terms plus boundary terms of the forms (a)~$\lim_{\Delta\rightarrow0^+} \int_{-\Delta}^{\Delta}\Theta(z)\delta(z)\dd{}z$ and (b)~$\lim_{\Delta\rightarrow0^+} \int_{-\Delta}^{\Delta}\Theta(z)\delta'(z)\dd{}z$, where $\Theta(z)$ is the Heaviside step function.
It is natural to assign the value $\tfrac{1}{2}$ to~(a).
As for terms~(b), which diverge, we must require that the sum of their prefactors vanishes, yielding a boundary condition that acts as a constraint in the minimization of $\omega_4$.

\section{\label{sec:semianalytical}Semianalytical solution}

\begin{figure}
  \includegraphics[scale=1.0]{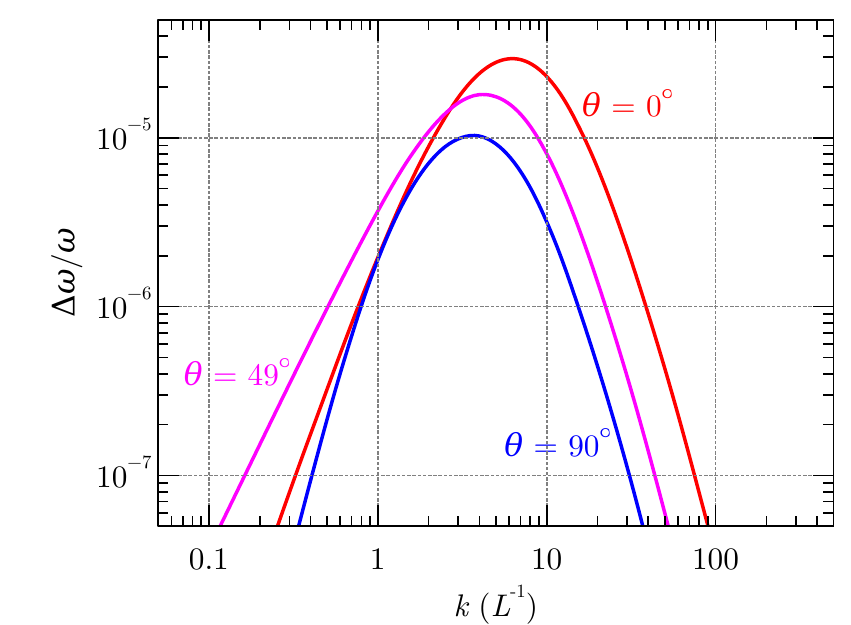}
  \caption{\label{fig:accuracy} (color online). Relative error $\Delta \omega / \omega$ of the semianalytical expression~\eqref{eq:blockgen} for the dispersion relation $\mathbf{\omega}(k\cos\vartheta,k\sin\vartheta)$, along three radials $\vartheta=0^\circ$ (region A), $\vartheta=90^\circ$ (region B), and $\vartheta=\vartheta_\text{cr}=49^\circ$ (A--B boundary), as compared to converged solutions of the full eigenvalue problem~\eqref{eq:modez} for $H_x = 0.73\,M_\text{S}$ and $2K = 0.46\,\mu_0 M_\text{S}^2$.
  In the interior of region~A, we have second-order accuracy in $k$ [$\Delta \omega = \Or(k^3)$]; for region~B, third-order accuracy [$\Delta \omega = \Or(k^4)$]; in region~C, third-order accuracy in $k^{-1}$ [$\Delta \omega = \Or(k^{-4})$].
  On the A--B boundary line, we have only first-order accuracy [$\Delta \omega = \Or(k^2)$].
  However, the relative error remains very small on the entire domain (well below $10^{-4}$).
  }
\end{figure}

In this section, we present a semianalytical expression for the dispersion relation of the $n=1$ \BVMSW{} mode that can be evaluated in constant time using standard numerical routines.
The expression is accurate up to an error that is negligible for any practical purpose (well below $0.01\%$ in the example of \figref{fig:accuracy}).
It takes a given wavevector $(k_x,k_y)$ as input; evaluation does not require an initial guess for $\omega$.

\begin{figure}
  \includegraphics[scale=1.0]{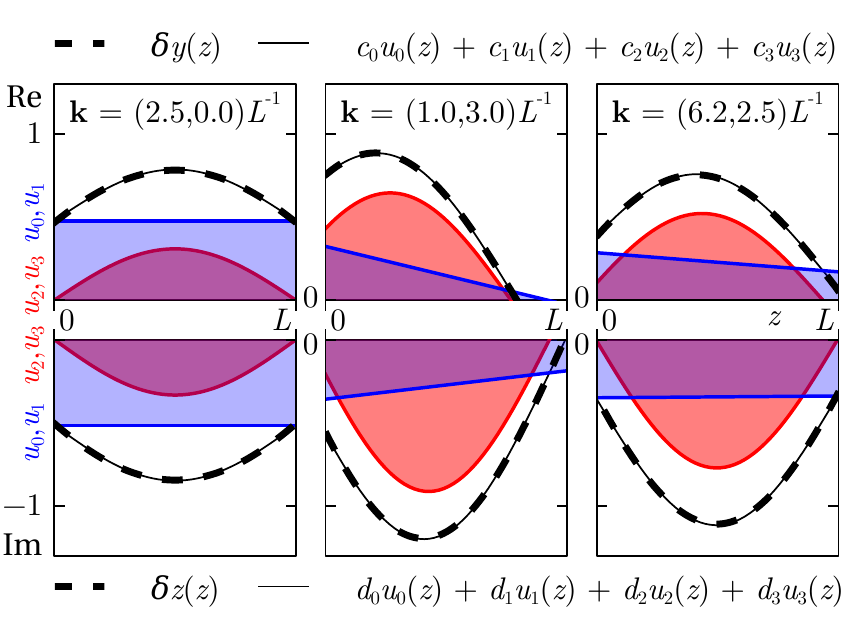}
  \caption{\label{fig:decomp} (color online).
  Mode profiles $\delta y(z), \delta z(z)$ of the lowest-frequency mode ($n=1$) and their approximate representations as linear combinations~\eqref{eq:deflincomb} of the basis functions $u_0, u_1, u_2, u_3$, for three wavevectors $\mathbf{k}=(k_x,k_y)$, taking $H_x = 0.73\,M_\text{S}$ and $2K = 0.46\,\mu_0 M_\text{S}^2$.
  There is no visible difference between the full profiles and the approximations.
  }
\end{figure}

\Figref{fig:decomp} shows that typical profiles $\delta y(z), \delta z(z)$ of the $n=1$ mode can be written, to a very reasonable approximation, as a linear combination of only four basis functions
\begin{subequations}\label{eq:basfuncs}
\begin{align}
u_0(z)& = \Pi^*\Bigl(\dfrac{z}{L}\Bigr) \text{,}\\
u_1(z)& = \Bigl(\dfrac{z}{L} - \dfrac{1}{2}\Bigr) \Pi^*\Bigl(\dfrac{z}{L}\Bigr) \text{,}\\
u_2(z)& = \sin\Bigl(\dfrac{\pi z}{L}\Bigr) \Pi^*\Bigl(\dfrac{z}{L}\Bigr) \text{,}\\
u_3(z)& = -\cos\Bigl(\dfrac{\pi z}{L}\Bigr) \Pi^*\Bigl(\dfrac{z}{L}\Bigr) \text{.}
\end{align}
\end{subequations}
For our semianalytical approximation, we restrict the profiles to such linear combinations
\begin{subequations}\label{eq:deflincomb}
\begin{align}
\delta y(z)& = c_0 u_0(z) + c_1 u_1(z) + c_2 u_2(z) + c_3 u_3(z)\text{,}\\
\delta z(z)& = d_0 u_0(z) + d_1 u_1(z) + d_2 u_2(z) + d_3 u_3(z)\text{.}
\end{align}
\end{subequations}
On this basis set, the operators $\hat{S}, \hat{D}^{yy}, \hat{D}^{yz}, \hat{D}^{zz}$ reduce to simple $4\times4$ matrix blocks, given below, and \equaref{eq:modez} becomes an $8\times8$ eigenvalue problem
\begin{multline}\label{eq:blockgen}
\left[
\begin{array}{c c}
H_x S + M_\text{S} D^{yy} & M_\text{S} D^{yz} \\
M_\text{S} D^{yz} & (H_x - \frac{2K}{\mu_0 M_\text{S}}) S + M_\text{S} D^{zz} \\
\end{array}
\right]
v
\\
= 
\frac{\omega}{\mu_0 |\gamma|}
\left[
\begin{array}{c c}
0 &  i  S \\
- i S & 0 \\
\end{array}
\right]
v
\text{,}
\end{multline}
where square brackets indicate block matrices and $v= (c_0,c_1,c_2,c_3,d_0,d_1,d_2,d_3)^T$ represents the eigenvector.
The approximate frequency of the $n=1$ mode is given by the lowest positive eigenvalue $\omega$.

\Equaref{eq:blockgen} takes the form of an $8\times8$ generalized Hermitian eigenvalue problem, the solutions $\omega$ of which can be found numerically using standard routines.
While not all linear-algebra computer packages may support the generalized format $Hv=\omega Qv$, it can always be rewritten as $Q^{-1} Hv=\omega v$ and solved as an ordinary non-Hermitian eigenvalue problem.

Explicit analytical expressions exist for all matrix elements in \equaref{eq:blockgen}. 
The identity operator $\hat{S}$ becomes the overlap matrix
\begin{equation}
S = L \left(
\begin{array}{c c c c}
1 & 0 & \tfrac{2}{\pi} & 0 \\
0 & \tfrac{1}{12} & 0 & \tfrac{2}{\pi^2} \\
\tfrac{2}{\pi} & 0 & \tfrac{1}{2} & 0 \\
0 & \tfrac{2}{\pi^2} & 0 & \tfrac{1}{2}\\
\end{array}
\right)
\text{,}
\end{equation}
where the matrix elements are defined by $S_{ij} = \int u_i(z) u_j(z) dz$.
The elements of the $D^{ab}$ matrix blocks can be evaluated in Fourier space
as
\begin{equation}
D^{yy}_{ij}(k_x,k_y) = \frac{1}{2\pi} \int \tilde{u}_i^*(k_z) \frac{k_y k_y}{k_x^2 + k_y^2 + k_z^2} \tilde{u}_j(k_z) dk_z
\text{,}
\end{equation}
and analogously for $D^{yz}_{ij}$ and $D^{zz}_{ij}$,
where $\tilde{u}_i(k_z)$ is the Fourier transform of the basis function $u_i(z)$.
We obtain
\begin{multline}
D^{yy} = L \frac{k_y^2}{k^2} \times \\
\left(
\begin{array}{c c c c}
1-N_{00} & 0 & \tfrac{2}{\pi}(1-N_{02}) & 0 \\
0 & \tfrac{1}{12}(1-N_{11}) & 0 & \tfrac{2}{\pi^2}(1-N_{13}) \\
\tfrac{2}{\pi}(1-N_{02}) & 0 & \tfrac{1}{2}(1-N_{22}) & 0 \\
0 & \tfrac{2}{\pi^2}(1-N_{13}) & 0 & \tfrac{1}{2}(1-N_{33})\\
\end{array}
\right)
\end{multline}
and
\begin{equation}
D^{zz} =
L \left(
\begin{array}{c c c c}
N_{00} & 0 & \tfrac{2}{\pi}N_{02} & 0 \\
0 & \tfrac{1}{12}N_{11} & 0 & \tfrac{2}{\pi^2}N_{13} \\
\tfrac{2}{\pi}N_{02} & 0 & \tfrac{1}{2}N_{22} & 0 \\
0 & \tfrac{2}{\pi^2}N_{13} & 0 & \tfrac{1}{2}N_{33}\\
\end{array}
\right)
\text{,}
\end{equation}
where
\begin{subequations}
\begin{align}
N_{00}& = \frac{1 - e^{-kL}}{kL}\text{,}\\
N_{11}& = 12\left[\frac{1 + e^{-kL}}{4 kL} + \frac{(1+kL)e^{-kL} -1}{k^3L^3}\right]\text{,}\\
N_{22}& = \frac{\pi^2}{k^2L^2 + \pi^2} - \frac{2 \pi^2 kL(1+e^{-kL})}{(k^2L^2 + \pi^2)^2}\text{,}\\
N_{33}& = \frac{\pi^2}{k^2L^2 + \pi^2} + \frac{2 k^3L^3 (1 + e^{-kL})}{(k^2L^2 + \pi^2)^2}\text{,}\\
N_{02}& = \frac{\pi^2}{k^2L^2 + \pi^2} \frac{1 + e^{-kL}}{2}\text{,}\\
N_{13}& = \frac{\pi^2}{k^2L^2 + \pi^2} \frac{(1+ e^{-kL})(2 + kL)}{4}
\end{align}
\end{subequations}
are the so-called demagnetizing factors;
and we obtain
\begin{equation}
D^{yz} = \frac{k_y L^2}{2\pi i}
\left(
\begin{array}{c c c c}
0 & \tfrac{\pi}{6} Z_{01} & 0 & \frac{2}{\pi} Z_{03} \\
- \tfrac{\pi}{6}Z_{01} & 0 & -\frac{4}{\pi^2}Z_{21} & 0 \\
0 & \frac{4}{\pi^2}Z_{21} & 0 & Z_{23} \\
-\frac{2}{\pi}Z_{03} & 0 & -Z_{23} & 0\\
\end{array}
\right)
\text{,}
\end{equation}
where
\begin{subequations}
\begin{align}
Z_{01}& = \frac{e^{-kL}+1 -2 N_{00}}{\tfrac{1}{6}k^2L^2}\text{,}\\
Z_{03}& = N_{02} \text{,}\\
Z_{21}& = \frac{\pi^2}{k^2L^2 + \pi^2} \left( 1 - \frac{\pi}{4}Z_{01} kL \right) \text{,}\\
Z_{23}& = N_{22} \text{.}
\end{align}
\end{subequations}

Notice that the basis set~\eqref{eq:basfuncs} has been chosen in such a way that it can represent exactly the profiles $\psi_0(z), \phi_1(z)$ of the $n=1$ mode in each of the regions A, B, and~C (see \tabref{tab:modeexpr}).
As a result, we have at least second-order accuracy of $\omega$ in $k$ or $1/k$ in those regions, as shown in \figref{fig:accuracy}. Since the mode profile $\psi_1(z)$ is fixed only at third (or, in regions~B and~C, fourth) order of perturbation theory (see \secref{sec:limbeh}), it does not need to be included in the basis set to obtain second-order accuracy.

On the boundary line between regions~A and~B ($\vartheta = \vartheta_\text{cr}$), we have only first-order accuracy, because the corresponding profiles $\psi_0(z), \phi_1(z)$ are not represented in the basis set.
\Figref{fig:accuracy} shows that the error $\Delta\omega$ nonetheless remains very small.
The exact small-$k$ behavior of $\omega$ on the A--B boundary is given in \secref{sec:regAB}.

We comment on our claim that the approximate expression for $\omega(k_x,k_y)$ can be evaluated in constant time.
In general, the solution of an eigenvalue problem for matrices of size $5\times 5$ or larger requires the use of iterative methods, the convergence rate of which may depend on system parameters.
In our case, the characteristic equation $\Det(H - \omega Q)=0$ of the eigenvalue problem~\eqref{eq:blockgen}
contains only even powers of $\omega$, since all eigenvalues appear in conjugate pairs (Hamiltonian problem \cite{Buijnsters2014}).
We could therefore write the characteristic equation, a polynomial of eighth degree in $\omega$, as a quartic polynomial in $\omega^2$, which can be explicitly solved by radicals.
This guarantees the existence of an analytical expression for $\omega$ in principle.

\section{\label{sec:conclusions}Conclusions}

\BVMSW{}s in magnetic films display unusual and highly nontrivial dispersion behavior.
Their strongly anisotropic and nonreciprocal propagation means that they can be excited and manipulated with a great deal of flexibility and control \cite{Satoh2012}.
Their specific dispersion characteristics are an important ingredient in the analysis of all-optical excitation \cite{Au2013,Shen2015} and nonlinear effects \cite{bauer1997direct, boyle1996nonlinear, Lennert2016}.

Since the defining equations of the \BVMSW{} modes can be solved only numerically, we believe that it is useful to have some approximate analytical results describing their essential features.
In \tabref{tab:asymp}, we summarize the simple analytical expressions that we have derived for the mode frequencies in the short-wavelength and long-wavelength regimes, including the behavior for wavevectors $\mathbf{k}$ pointing in a direction close to the critical angle $\vartheta_\text{cr}$.
We have also obtained explicit first-order expressions for the depth profiles $\delta y(z),\delta z(z)$ of the modes, given by \equarefs{eq:genform}, \eqref{eq:expandsmallk}, \eqref{eq:expandbigk}, and \tabref{tab:modeexpr}. These expressions highlight and quantify the asymmetry in $z$ found for $k_y \neq 0$ (nonreciprocal behavior).

In addition to the perturbative results, we provide a semianalytical expression for the dispersion relation of the lowest mode $n=1$ valid for arbitrary wavevector $(k_x,k_y)$. While this expression is, strictly speaking, an approximation, we find that the error is so small as to be negligible for practical purposes.
The semianalytical expression is straightforward to implement using standard numerical routines.

\appendix

\section{\label{sec:distlimit}Distributional limits}

This appendix provides some elementary results needed to carry out the
small-$k$ expansion~\eqref{eq:Dexpsmall}.
If we set $(k_x,k_y) = (k\cos\vartheta, k\sin\vartheta)$, the operators~$\hat{D}^{yy}, \hat{D}^{yz}, \hat{D}^{zz}$, defined by \equaref{eq:defD}, become
\begin{subequations}
\begin{align}
\hat{D}^{yy}& = \frac{k^2 }{k^2 + \hat{k}_z^2} \sin^2\vartheta \text{,} \\
\hat{D}^{yz}& = \frac{k \hat{k}_z }{k^2 + \hat{k}_z^2} \sin\vartheta\text{,} \\
\hat{D}^{zz}& = \frac{\hat{k}_z^2}{k^2 + \hat{k}_z^2} = \hat{S} - \frac{k^2}{k^2 + \hat{k}_z^2}\text{.}
\end{align}
\end{subequations}
\Equaref{eq:Dexpsmall} is obtained by expanding the operators $k^2 / (k^2 + \hat{k}_z^2)$ and $k \hat{k}_z / (k^2 + \hat{k}_z^2)$ as a Taylor series in the parameter $k>0$.
In the following, $\hat{k}_z$ may be substituted for $x$ and $k$ for $\varepsilon$.

We consider the expressions $\varepsilon^2 / (x^2 + \varepsilon^2)$ and $\varepsilon x / (x^2 + \varepsilon^2)$
as functions of $x$ and calculate derivatives with respect to the parameter $\varepsilon$ in the limit $\varepsilon\rightarrow0^+$.
Most of these limits can only be defined if we turn to generalized functions (distributions) of~$x$.
We obtain
\begin{subequations}
\begin{align}
\lim_{\varepsilon\rightarrow0^+} \frac{\varepsilon^2}{x^2 + \varepsilon^2}& = 0\text{,} \\
\lim_{\varepsilon\rightarrow0^+} \frac{\varepsilon x}{x^2 + \varepsilon^2}& = 0\text{;}
\end{align}
\end{subequations}
\begin{subequations}
\begin{align}
\lim_{\varepsilon\rightarrow0^+} \frac{\partial}{\partial\varepsilon} \frac{\varepsilon^2}{x^2 + \varepsilon^2}& = \pi \delta(x)\text{,} \\
\lim_{\varepsilon\rightarrow0^+} \frac{\partial}{\partial\varepsilon} \frac{\varepsilon x}{x^2 + \varepsilon^2}& =
\frac{1}{x}\text{,}
\end{align}
\end{subequations}
where $\delta(x)$ is the Dirac delta distribution;
\begin{subequations}
\begin{align}
\lim_{\varepsilon\rightarrow0^+} \frac{\partial^2}{\partial\varepsilon^2} \frac{\varepsilon^2}{x^2 + \varepsilon^2}& = 
\frac{2}{x^2}\text{,} \\
\lim_{\varepsilon\rightarrow0^+} \frac{\partial^2}{\partial\varepsilon^2} \frac{\varepsilon x}{x^2 + \varepsilon^2}& = 2\pi \delta'(x) \text{;}
\end{align}
\end{subequations}
and
\begin{subequations}
\begin{align}
\lim_{\varepsilon\rightarrow0^+} \frac{\partial^3}{\partial\varepsilon^3} \frac{\varepsilon^2}{x^2 + \varepsilon^2}& = -3\pi \delta''(x)\text{,} \\
\lim_{\varepsilon\rightarrow0^+} \frac{\partial^3}{\partial\varepsilon^3} \frac{\varepsilon x}{x^2 + \varepsilon^2}& = -\frac{6}{x^3}
\text{.}
\end{align}
\end{subequations}
Expressions of the form $1/x^n$ are formally defined as distributional derivatives $ \frac{(-1)^{n-1}}{(n-1)!} \frac{\ddnoskip{}^n}{{\ddnoskip{}x}^n} \log |x|$.

\section{\label{sec:pertthy}Perturbation theory}

The generalized Hermitian eigenvalue problem
\begin{equation}
H\Psi = \omega Q \Psi\text{,}
\end{equation}
where $H$ and $Q$ are Hermitian operators one of which is positive definite,
can be cast as a problem of minimization of the functional
\begin{equation}\label{eq:verygeneralH}
\omega = \Psi^\dagger H \Psi
\end{equation}
under the constraint
\begin{equation}\label{eq:verygeneralQ}
\Psi^\dagger Q \Psi = 1 \text{.}
\end{equation}
We suppose that $H$ depends on a parameter $k$ and expand the solution $\Psi$ around $k=0$.
\Equarefs{eq:verygeneralH} and~\eqref{eq:verygeneralQ} become
\begin{multline}
(\omega_0 + k\omega_1 + \ldots) = (\Psi_0 + k\Psi_1 + \ldots)^\dagger \\
 \cdot (H_0 + k H_1 + \ldots) (\Psi_0 + k\Psi_1 + \ldots)
\end{multline}
and
\begin{equation}
(\Psi_0 + k\Psi_1 + \ldots)^\dagger Q (\Psi_0 + k\Psi_1 + \ldots) = 1
\text{.}
\end{equation}
Collecting like powers of $k$, we obtain
\begin{subequations}\label{eq:genfunc}
\begin{align}
\omega_0& = \Psi_0^\dagger H_0 \Psi_0 \text{,} \\
\label{eq:func1gen}\omega_1& = 2 \Psi_0^\dagger H_0 \Psi_1 + \Psi_0^\dagger H_1 \Psi_0\text{,} \\
\omega_2& = 2 \Psi_0^\dagger H_0 \Psi_2 + \Psi_1^\dagger H_0 \Psi_1 \nonumber \\
 & \phantom{={}} + 2\Psi_0^\dagger H_1 \Psi_1 + \Psi_0^\dagger H_2 \Psi_0\text{,} \\
\omega_3& = 
  2\Psi_0^\dagger H_0 \Psi_3 + 2\Psi_1^\dagger H_0 \Psi_2 \nonumber\\
 & \phantom{={}}  + 2\Psi_0^\dagger H_1 \Psi_2 + \Psi_1^\dagger H_1 \Psi_1 \nonumber\\
 & \phantom{={}} + 2\Psi_0^\dagger H_2 \Psi_1 + \Psi_0^\dagger H_3 \Psi_0\text{,} \\
\omega_4& = 
  2 \Psi_0^\dagger H_0 \Psi_4 + 2 \Psi_1^\dagger H_0 \Psi_3 + \Psi_2^\dagger H_0 \Psi_2 \nonumber\\
 & \phantom{={}} + 2\Psi_0^\dagger H_1 \Psi_3 + 2\Psi_1^\dagger H_1 \Psi_2 + 2\Psi_0^\dagger H_2 \Psi_2 \nonumber\\
 & \phantom{={}} + \Psi_1^\dagger H_1 \Psi_1 + 2\Psi_0^\dagger H_3 \Psi_1 + \Psi_0^\dagger H_4 \Psi_0
\end{align}
\end{subequations}
and
\begin{subequations}\label{eq:genconstr}
\begin{align}
1& = \Psi_0^\dagger Q \Psi_0 \text{,} \\
\label{eq:constr1gen}0& = 2 \Psi_0^\dagger Q \Psi_1 \text{,} \\
0& = 2 \Psi_0^\dagger Q \Psi_2 + \Psi_1^\dagger Q \Psi_1 \text{,}\\
0& = 2 \Psi_0^\dagger Q \Psi_3 + 2 \Psi_1^\dagger Q \Psi_2 \text{,}\\
0& = 2 \Psi_0^\dagger Q \Psi_4 + 2\Psi_1^\dagger Q \Psi_3 + \Psi_2^\dagger Q \Psi_2 \text{,}
\end{align}
\end{subequations}
where we assume that all terms $\Psi_A^\dagger H \Psi_B$ and $\Psi_A^\dagger Q \Psi_B$ are real.
To obtain the expansion $\Psi_0 + k\Psi_1 + k^2\Psi_2 + \ldots$ of the eigenfunction $n=1$ with the lowest eigenvalue $\omega$, we sequentially minimize the functionals $\omega_0, \omega_1, \omega_2, \ldots$ under the constraints~\eqref{eq:genconstr}.

In the following, we assume that the eigenvectors $\Psi$ are written in the form~\eqref{eq:genform}, and we assume that $Q$ and $H_0$ are given by
\begin{align}
 \Psi_A^\dagger Q \Psi_B& = \psi_A \psi_B - \phi_A \phi_B \text{,} \\
 \Psi_A^\dagger H_0  \Psi_B& = \mu_0 |\gamma| a b \bigl(\psi_A \psi_B + \phi_A \phi_B\bigr)\text{.}
\end{align}
The positive-$\omega$ solutions of the zeroth-order eigenvalue equation $H_0\Psi_0 = \omega_0 Q \Psi_0$ have $\phi_0=0$ and $\psi_0$ arbitrary (provided $\psi_0^2=1$).
We further assume that the $H_i$ for $i\ge 1$ are of the form
\begin{equation}
H_i =
\left( 
\begin{array}{c c}
\hat{A}_i & \hat{C}_i \\
\hat{C}_i & \hat{B}_i
\end{array}
\right)
\text{,}
\end{equation}
where $\hat{C}$ is Hermitian. We have
\begin{equation}
\Psi_A^\dagger H_i  \Psi_B = \psi_A \hat{X}_i \psi_B + \phi_A \hat{X}_i \phi_B + \psi_A \hat{Y}_i \phi_B + \phi_A \hat{Y}_i^\dagger \psi_B
\end{equation}
with
\begin{subequations}
\begin{align}
\hat{X}_i& = \frac{\mu_0 |\gamma|}{2ab} \bigl(a^2 \hat{A}_i + b^2 \hat{B}_i\bigr) \text{,} \\
\hat{Y}_i& = \frac{\mu_0 |\gamma|}{2ab} \bigl(-a^2 \hat{A}_i + b^2 \hat{B}_i - 2iab \hat{C}_i \bigr) \text{.}
\end{align}
\end{subequations}
The functionals~\eqref{eq:genfunc} become
\begin{subequations}
\begin{align}
\omega_0& = \mu_0|\gamma|ab\text{,} \\
\label{eq:func1neat}\omega_1& = \psi_0 \hat{X}_1 \psi_0\text{,}\\
\label{eq:func2neat}\omega_2& = 2\omega_0\phi_1^2 + 2\psi_0\hat{X}_1\psi_1 + 2\psi_0\hat{Y}_1\phi_1 + \psi_0\hat{X}_2\psi_0\text{,} \\
\omega_3& = 4\omega_0\phi_1\phi_2 + 2 \psi_0 \hat{Y}_1\phi_2 \nonumber\\
 &\phantom{={}} + \psi_1 \hat{X}_1 \psi_1 + \phi_1 \hat{X}_1 \phi_1 + 2\psi_1 \hat{Y}_1 \phi_1  \nonumber\\
 &\phantom{={}} + 2\psi_0 \hat{X}_1 \psi_2 + 2\psi_0 \hat{X}_2 \psi_1 + 2 \psi_0 \hat{Y}_2\phi_1 \nonumber\\
 &\phantom{={}} + \psi_0 \hat{X}_3 \psi_0
\text{,}\\
\omega_4& = 4\omega_0 \phi_1\phi_3 + 2\omega_0\phi_2^2 \nonumber\\
 &\phantom{={}} + 2\psi_0 \hat{X}_1 \psi_3 + 2\psi_0 \hat{Y}_1 \phi_3 + 2\psi_1\hat{X}_1\psi_2 + 2\phi_1\hat{X}_1\phi_2 \nonumber\\
 &\phantom{={}} + 2\psi_1\hat{Y}_1\phi_2 + 2\phi_1\hat{Y}_1^\dagger\psi_2 \nonumber\\
 &\phantom{={}} + 2\psi_0\hat{X}_2\psi_2 + 2\psi_0\hat{Y}_2\phi_2 \nonumber\\
 &\phantom{={}} + \psi_1 \hat{X}_2\psi_1 + \phi_1 \hat{X}_2\phi_1 + 2\phi_1 \hat{Y}_2\psi_1 \nonumber\\
 &\phantom{={}} + 2\psi_0\hat{X}_3\psi_1 + 2\psi_0\hat{Y}_3\phi_1 + \psi_0\hat{X}_4\psi_0
\text{,}
\end{align}
\end{subequations}
where we have substituted constraint~\eqref{eq:constr1gen} into \equaref{eq:func1gen} and so on.
The constraints~\eqref{eq:genconstr} become
\begin{subequations}
\begin{align}
\label{eq:constr0neat}1 &= \psi_0^2\text{,}\\
\label{eq:constr1neat}0 &= 2 \psi_0 \psi_1\text{,}\\
\label{eq:constr2neat}0 &= 2 \psi_0 \psi_2 + \psi_1^2 - \phi_1^2\text{,}\\
0 &= 2\psi_0\psi_3 + 2\psi_1\psi_2 - 2\phi_1\phi_2 \text{.}
\end{align}
\end{subequations}
If we wish to obtain the higher modes $n=2,3,\ldots$, we carry out the minimization of the functionals $\omega_i$ under additional constraints $\psi_0^{(m)}\psi_0^{(n)}=0$, $\psi_0^{(m)}\psi_1^{(n)}+\psi_0^{(n)}\psi_1^{(m)}=0$, etc.\ for all $m<n$.

The function $\psi_0$ is found by minimization of $\omega_1$~\eqref{eq:func1neat}, lifting the degeneracy at zeroth order. It satisfies
\begin{equation}
\hat{X}_1 \psi_0 = \omega_1 \psi_0\text{.}
\end{equation}
Together with \equarefs{eq:constr1neat} and~\eqref{eq:constr2neat}, this implies $\psi_0\hat{X}_1\psi_1 = 0$ and $2 \psi_0 \hat{X}_1 \psi_2 = - \omega_1 \bigl[ \psi_1^2 - \phi_1^2 \bigr]$.
Given $\psi_0$, minimization of $\omega_2$~\eqref{eq:func2neat} then yields
\begin{equation}\label{eq:phi1sol}
\phi_1 = -\tfrac{1}{2\omega_0} \hat{Y}_1^\dagger \psi_0\text{.}
\end{equation}
If $\psi_0$ is not yet (completely) determined by minimization of $\omega_1$, it may be obtained by minimization of
\begin{equation}\label{eq:phi0func2nd}
\omega_2 = \psi_0\hat{X}_2\psi_0 -\frac{1}{2\omega_0} \psi_0 \hat{Y}_1 \hat{Y}_1^\dagger \psi_0 \text{,}
\end{equation}
where we have substituted \equaref{eq:phi1sol} into \equaref{eq:func2neat}.

Notice that, owing to the degeneracy of the modes at zeroth order, the other first-order component $\psi_1$ is not fixed by minimization of $\omega_2$.
We minimize
\begin{align}
\omega_3& = \psi_1 (\hat{X}_1 - \omega_1) \psi_1 + \phi_1 (\hat{X}_1+\omega_1) \phi_1  \nonumber\\
 &\phantom{={}}  + 2\psi_1 \hat{Y}_1 \phi_1  + 2\psi_0 \hat{X}_2 \psi_1 + 2 \psi_0 \hat{Y}_2\phi_1 \nonumber\\
\label{eq:om3func} &\phantom{={}} + \psi_0 \hat{X}_3 \psi_0
\text{,}
\end{align}
yielding the condition
\begin{equation}\label{eq:psi1sol}
\bigl( \hat{X}_1 - \omega_1 \bigr) \psi_1 = - \hat{Y}_1 \phi_1 -\hat{X}_2 \psi_0 + \lambda \psi_0\text{,}
\end{equation}
where $\lambda$ is chosen such that the equation has a solution. The solution $\psi_1$ is now defined up to a term $\propto \psi_0$, which is fixed by the constraint~\eqref{eq:constr1neat}.

If $ \hat{X}_1 - \omega_1 $ has null vectors other than $\psi_0$, the profile $\psi_1$ is not yet (completely) fixed by the condition~\eqref{eq:psi1sol}. We then consider
\begin{align}
\omega_4& = 2\omega_0\phi_2^2 \nonumber\\
 &\phantom{={}} + 2\bigl[\bigl(\hat{X}_1+\omega_1\bigr)\phi_1 + \hat{Y}_1^\dagger \psi_1 + \hat{Y}_2^\dagger \psi_0\bigr]^\dagger\phi_2 \nonumber\\
 &\phantom{={}} + \psi_1 \bigl(\hat{X}_2-\lambda\bigr)\psi_1 + \phi_1 \bigl(\hat{X}_2 + \lambda\bigr)\phi_1 + 2 \psi_1 \hat{Y}_2\phi_1 \nonumber\\
 &\phantom{={}} + 2\psi_0\hat{X}_3\psi_1 + 2\psi_0\hat{Y}_3\phi_1 + \psi_0\hat{X}_4\psi_0
\text{,}
\end{align}
where $\lambda = \psi_0 \hat{Y}_1 \phi_1 + \psi_0 \hat{X}_2 \psi_0$.
Minimization gives
\begin{equation}
\phi_2 = \frac{-1}{2\omega_0}\bigl[
   \bigl(\hat{X}_1+\omega_1\bigr)\phi_1 + \hat{Y}_1^\dagger \psi_1 + \hat{Y}_2^\dagger \psi_0 \bigr]
\text{.}
\end{equation}
Eliminating $\phi_2$, we obtain
\begin{align}
\omega_4& = -\frac{1}{2\omega_0} \bigl\lVert \bigl(\hat{X}_1+\omega_1\bigr)\phi_1 + \hat{Y}_1^\dagger \psi_1 + \hat{Y}_2^\dagger \psi_0 \bigr\rVert^2 \nonumber\\
 &\phantom{={}} + \psi_1 \bigl(\hat{X}_2-\lambda\bigr)\psi_1 + \phi_1 \bigl(\hat{X}_2 + \lambda\bigr)\phi_1 + 2 \psi_1 \hat{Y}_2\phi_1 \nonumber\\
\label{eq:om4func} &\phantom{={}} + 2\psi_0\hat{X}_3\psi_1 + 2\psi_0\hat{Y}_3\phi_1 + \psi_0\hat{X}_4\psi_0
\text{,}
\end{align}
which functional should be minimized treating \equaref{eq:psi1sol} as an additional constraint.

\end{document}